\documentclass[a4paper]{article}

\usepackage[a4paper,margin=2.5cm]{geometry}

\usepackage{bussproofs,stackengine,graphicx}
\usepackage{tikz}
\usetikzlibrary{arrows.meta, positioning}
\usepackage{mathdots}
\usepackage{todonotes}
\usepackage{amsthm, amsmath, amsfonts, amssymb,bm}
\usepackage[ruled,vlined,algosection,norelsize]{algorithm2e}
\usepackage{booktabs}
\usepackage{float}

\newcommand{\eps}{\epsilon}
\newcommand{\bb}[1]{\mathbb{#1}}
\newcommand{\bff}[1]{\mathbf{#1}}
\newcommand{\ms}[1]{\mathsf{#1}}

\newcommand{\comp}[1]{\sf{Comp(#1)}}

\newcommand{\BSI}{\ms{Conc}\text{-}\ms{I}_\ms{W}}
\newcommand{\SSI}{\ms{Sum}\text{-}\ms{I}_\ms{W}}
\newcommand{\SSIl}{\ms{Sum}\text{-}\ms{I}_\ms{\ell}}

\newcommand{\To}{\Rightarrow}
\newcommand{\rhdd}{\blacktriangleright}
\newcommand{\pair}[1]{\langle{#1}\rangle}

\newcommand{\fun}[1]{\mathsf{#1}}

\newcommand{\cc}[1]{\mathbf{#1}}

\newcommand{\FP}{\cc{FP}}
\newcommand{\AC}{\cc{FAC}}

\newcommand{\PH}{\cc{PH}}
\newcommand{\NP}{\cc{NP}}
\newcommand{\coNP}{\cc{coNP}}

\newcommand{\sP}{\cc{\#P}}
\newcommand{\PP}{\cc{PP}}
\newcommand{\FCH}{\cc{FCH}}

\newcommand{\Nat}{\mathbb{N}}

\newcommand{\acc}{\mathtt{acc}}

\newcommand{\prob}[1]{\mathtt{#1}}

\newcommand{\WProd}{\fun{WProd}}
\newcommand{\Sum}{\fun{Sum}}

\newcommand{\remov}[1]{}

\usepackage{hyperref}
\theoremstyle{definition}
\newtheorem{theorem}{Theorem}[section]
\newtheorem{definition}[theorem]{Definition}
\newtheorem{proposition}[theorem]{Proposition}
\newtheorem{remark}[theorem]{Remark}
\newtheorem{lemma}[theorem]{Lemma}

\newtheorem{notation}[theorem]{Notation}
\title{Applicative theories can count}

\author{Melissa Antonelli$^1$ \and Marcel Ertel$^1$ \and Eduardo Skapinakis$^{1,2}$}
\date{$^1$Carl Friedrich von Weizs{\"a}cker Zentrum, Universit{\"a}t T{\"u}bingen\\
$^2$Center for Mathematics and Applications (NOVA Math), NOVA FCT\\
\vspace{3pt}\today}

\begin{document}

\maketitle

\begin{abstract}
Applicative theories offer a proof-theoretic framework for studying complexity. In this paper, we introduce a family of applicative theories to characterize classes ranging from constant-depth circuits to polynomial space. Our approach relies on a simple axiom scheme, the small variations of which allows to capture various forms of counting and alternation.
This leads to highly uniform results. When applied on notation, the scheme captures bit counting, yielding a characterization of the functions computable by constant-depth, polynomial-size threshold circuits ($\bff{FTC}^0$); when applied lexicographically, it allows for the counting of accepting computations, yielding a characterization of the functions in the counting hierarchy ($\bff{FCH}$).
Increasing the logical complexity of the induction formulas allowed in the scheme, we obtain characterizations of the functions computable by bounded Boolean circuits of logarithmic-depth and polynomial-size ($\bff{FNC}^1$) and of the functions computable in polynomial space ($\bff{FPSPACE}$). 
The latter characterization solves an open problem in the field.
Finally, we investigate separation results. 
While it is unknown whether $\bff{FNC}^1$ and $\bff{FCH}$ are different, there is an oracle relative to which (relativized) $\bff{FNC}^1$ is strictly contained in (relativized) $\bff{FCH}$. We use this fact to separate the theories characterizing them, by exploring their ability to represent higher-order functionals. 
\end{abstract}

\newpage

\section{Introduction}\label{sec:intro}
First introduced by Feferman in the 1970s to provide logical foundations for Bishop-style constructive mathematics~\cite{feferman:75,feferman:79}, applicative theories have subsequently been used for studying computational complexity~\cite{cantini:02,strahm:03} (see~\cite{strahm:10} for a survey).
Since their language is expressive enough to define terms representing any partial recursive function, these theories offer a convenient framework for obtaining proof-theoretic characterizations of classes, without coding functions as complex formulas.
The intuitive idea is that one can reinterpret an equation of the form $F(x+1) = H(F(x))$ as the implication $F(x) = z \to F(x+1) = H(z)$, which, letting $A(x,z)$ be the atomic formula $F(x)=z$, can be seen as the induction step $A(x,z) \to A(x+1,H(z))$.
The advantages of the applicative-theoretic perspective over other machine-independent approaches to complexity are (at least) twofold:
first, induction schemes offer more degrees of freedom than recursion schemes, namely, the logical complexity of the formulas allowed in induction and the scope of the quantifiers involved~\cite{strahm:03,eberhard:15};
second, applicative theories can elegantly capture higher-order versions of complexity classes~\cite{strahm:04}, thus providing a uniform setting for studying how the computational content of certain induction schemes propagates to proofs of complex formulas.
Following the initial work devoted to poly-time computable functions~\cite{strahm:97,cantini:02,strahm:03}, multiple complexity classes have been characterized relying on restricted principles of induction~\cite{cantini:02,strahm:03,calamai:08,kahle:11,eberhard:15}.
However, thus far, small-circuit and counting classes have mostly escaped applicative-theoretic treatment.

In this paper, we bridge this gap by introducing and studying \emph{uniform} applicative theories to characterize function classes ranging from small circuit classes ($\cc{FAC}^0$) to polynomial space ($\cc{FPSPACE}$). We always consider the circuit classes with $\bff{DLOGTIME}$-uniformity \cite{ruzzo:81,barrington:90}.
Technically, these results are achieved by considering specific variations of the main inductive schemes.
To deal with \textit{counting} classes~\cite{valiant:79,wagner:86,toran:91}, we introduce a new scheme designed to straightforwardly support operations related to enumerating and counting. This leads to the very first applicative-theoretic characterization of functions computed by constant-depth threshold circuits ($\cc{FTC}^0$) and of the counting hierarchy ($\cc{FCH}$).
Remarkably, the difference between the schemes characterizing these two classes lies solely in whether our original ``Sum induction'' axiom $(\ms{Sum}\text{-}\ms{I})$ is formulated on notation, i.e.,~$A(x)\to A(xi)$, or lexicographically, i.e.,~$A(x)\to A(\ms{s_\ell}x)$.
Additionally, we show that, by introducing disjunctions in the induction formula, one can simulate forms of constant-bounded recursion~\cite{clote:99}, thereby characterizing the functions in $\bff{FPSPACE}$ as well as those computed by poly-sized Boolean circuits of bounded fan-in in logarithmic depth ($\bff{FNC}^1$).
The former characterization solves an open problem posed in~\cite{kahle:11}, while the latter recalls a similar result of~\cite{eberhard:15}.
Overall, this work shows that simple intuitions in the applicative setting offer a new framework to study and investigate counting and alternating classes in terms of syntactical constraints on induction schemes.
Our results are summarized in Table~\ref{tab:theories}.

\begin{table}[H]
\centering
\small
\begin{tabular}{c c c c}
\toprule
\textbf{Theory} & \textbf{Axiom Scheme} & \textbf{Complexity Class} & \textbf{Proof} \\
\midrule
${\sf LCHT}$  &   ${\sf LHT} + (\SSI)$        & \;  $\bff{FTC}^0$      &  Prop.~\ref{prop:lower_bound_counting_TC}, \ref{prop:upper_bound_tc0}   \\
${\sf ALT}$  & \;   ${\sf LHT} + (\SSI)^\vee$   & \;  $\bff{FNC}^1$      &  Prop.~\ref{prop:lower_bound_nc1}, \ref{prop:upper_bound_nc1}    \\
${\sf CHT}$   &   ${\sf LHT} + (\SSIl)$       &  $\bff{FCH}$        &  Prop.~\ref{prop:lower_bound_counting}, \ref{prop:upper_bound_fch}   \\
${\sf PST}$   &   ${\sf LHT} + (\SSIl)^\vee$  & $\bff{FPSPACE}$   &  Prop.~\ref{prop:lower_bound_pspace}, \ref{prop:upper_bound_pspace}    \\
\bottomrule
\end{tabular}
\caption{Applicative theories and corresponding characterizations}\label{tab:theories}
\end{table}

\normalsize


\noindent
This picture also illustrates how computational resources are affected (or unaffected) by modifications in the logical machinery corresponding to the induction scheme that captures a given form of recursion (see~\cite{strahm:03,eberhard:15} for similar considerations).

Some of the classes we study are known to be distinct. Namely, since we consider $\bff{DLOGTIME}$-uniform circuit families, $\mathbf{TC}^0\subsetneq \mathbf{CH}$ \cite{caussinus:98,allender:99} and $\bff{NC}^1\subsetneq\bff{PSPACE}$ (via $\bff{NC}^1\subseteq\bff{L}$ \cite{borodin:77} and $\bff{L}\subsetneq\bff{PSPACE}$ \cite{stearns:65}).
The other relationships between these classes are open.
Some even remain open in the relativized setting: it is not known whether there is an oracle separating $\bff{FCH}$ from $\bff{PSPACE}$, nor $\bff{FTC}^0$ from $\mathbf{FNC}^1$. Indeed, the latter is equivalent to an unrelativized separation~\cite{aehlig:16}.
Relying on the fact that there is an oracle separating $\bff{FNC}^1$ from $\bff{FCH}$, we show that the corresponding applicative theories are different, i.e.,~${\sf ALT}\neq {\sf CHT}$, yielding the first separation result for such weak systems.
This separation is based on the fact that applicative theories can express higher-order functionals, and thus totality statements about oracle Turing machines.

\bigskip

\noindent
\textbf{Structure of the paper.}
Sec.~\ref{sec:background} recalls standard preliminary notions from circuit and counting complexity, and the basics of applicative theories. 
Sec.~\ref{sec:the_theories} introduces a family of original induction schemes to define corresponding uniform applicative theories. 
Secs.~\ref{sec:lowerB} and \ref{sec:upper_bounds} are devoted to showing that these theories effectively capture the desired complexity classes, proving the main characterization results by establishing the lower and upper bounds for each theory.
Sec.~\ref{sec:separatin_oracle} establishes relativized separation between $\bff{FNC}^1$ and $\bff{FCH}$ via higher-order functionals. 
Sec.~\ref{sec:conclusion} concludes the paper with some open directions for future research.
\section{Background and notation}\label{sec:background}
We assume that the reader is familiar with standard notions in complexity, proof theory, and applicative theories.
Here, we recall the main notions in these areas and make our notational conventions explicit.
For comprehensive presentations, see~\cite{balcazar:88,clote:99}, \cite{strahm:03,cantini:05,strahm:10}, and~\cite{troelstra:00}.

\bigskip

\noindent
\textbf{On circuit and counting complexity.}
We start by fixing notational conventions.

\begin{notation}\label{not:notation}
Let $\bb{W} = \{0,1\}^*$ be the set of 0-1 words, where $\epsilon$ is the empty word. 
For any $x\in \bb{W}$, let $|x|$ denote the number of its bits, and $x(i)$ indicate its $i^{th}$ bit: $x=x(|x|-1)\dots x(0)$.
\end{notation}

\noindent
The set $\bb{W}$ is in one-to-one correspondence with the binary representation of natural numbers, following the ordering $\eps, 0, 1, 00, 01, 10, 11, 000,\dots$.
\normalsize
The classes (of functions) considered in this work are defined in the standard way. To keep the presentation of counting classes as compact as possible, we follow the approach that uses the counting operator $\mathbb{C}$~\cite{wagner:86,allender:90}.
Recall also that the pairing operation $\pair{\cdot,\cdot}$ on $\bb{W}$ is defined as in~\cite{clote:99}, so that it and its associated projection functions $(\cdot)_0$ and $(\cdot)_1$ are in $\bff{FAC}^0$.

\begin{definition}\label{def:classes}
We use $\FP$ (resp.,~$\bff{FPSPACE}$) to denote the class of functions from (the Cartesian products of) $\bb{W}$ to $\bb{W}$, which are computable by deterministic Turing machines (TM) in time (resp.,~space) polynomial in the size of the input.
\\
For any class of languages, $\bff{C}$, we use $\mathbb{C} \cdot \bff{C}$ to denote the set of languages $L$ for which there is a polynomial $p$ and a set $A\in \bff{C}$ such that:
$ 
x \in L$
iff
 $|\{ y : |y| \leq p(|x|) \text{ and } \langle x,y\rangle \in A\}| \geq 2^{p(|x|)}$.
Then, the counting hierarchy ($\bff{CH}$)~\cite{wagner:86} is defined inductively: 
$
\bff{CH}_0 = \cc{P}, 
\bff{CH}_{i+1} = \bb{C} \cdot \bff{CH}_i$ 
and $\bff{CH} = \bigcup_{i\ge 0} \bff{CH}_i$.
We use $\bff{FCH}$ for the function version of the hierarchy~\cite{vollmer:96}.
\\
For $i\in \Nat$, $\mathbf{AC}^i$ (resp.,~$\mathbf{NC}^i$) is the class of languages recognized by a $\mathbf{DLOGTIME}$-uniform (see \cite{ruzzo:81,barrington:90}) family of unbounded (resp.,~bounded) fan-in Boolean circuits of polynomial size and depth $O((\log n)^i)$.
The basic level of this hierarchy is $\mathbf{AC}^0$, defined by unbounded circuits of polynomial size and constant depth.
If  circuits also include majority (or threshold) gates, the class is called $\mathbf{TC}^i$.
We use $\mathbf{FAC}^i$ (resp., $\bff{FNC}^i$) and $\mathbf{FTC}^i$ to denote the corresponding function classes.
\end{definition}

\textbf{Applicative theories.}
The applicative theories we are going to consider are all defined over the first-order language $\mathcal{L}$, including individual constant symbols corresponding to: the combinators $\ms{k}$ and $\ms{s}$; the empty word ($\epsilon$); the binary successors, concatenating the given input with either 0 or 1  ($\ms{s_0}$, $\ms{s_1}$); the binary predecessor, removing the right-most bit of the input ($\ms{p_w}$); lexicographic successor and predecessor, resp., adding and subtracting 1 from the input according to the ordering given above ($\ms{s_\ell}$, $\ms{p}_\ell$); word concatenation ($\ms{*}$); word multiplication, concatenating the first input with itself the size of the second input many times ($\ms{\times}$); conditional on words, returning a different word depending on whether its first input is $\epsilon$, ends in 0 or ends in 1 $(\ms{c_W}$); the length symbol, returning the length of a word ($\ms{len}$); the bit symbol, returning a given bit of $x$ ($\ms{bit}$); the eraser symbol ($\ms{e}$), returning the input with all its right-most 0's removed.
Additionally, $\cal{L}$ contains a set of variables and its set of terms is closed under term application, denoted by $\cdot$, which gives ``applicative theories'' their name.

\begin{definition}
    Terms in the language $\mathcal{L}$ are defined by the following syntax:
$$
    t:=  \mathcal{X} \;| \;\ms{k}\;|\; \ms{s}\;|\; \epsilon\; |\; \ms{s_0}\;|\; \ms{s_1}\;|\; \ms{p_w}\;|\; \ms{s_\ell}\; |\; \ms{p_\ell}\;|\;\ms{len}\;|\; \ms{*}\;|\; \ms{\times}\;|\; \ms{c_W}\;|\; \ms{bit}\;|\; \ms{e}\;| \; (t \cdot t)
$$
where $x\in\mathcal{X}$, being $\mathcal{X} = \{x,y,z,f,g,h,\dots\}$ a set of variables (possibly containing subscripts).
The grammar for formulas of $\mathcal{L}$ is given below:
$$
    A:=\ms{W}(t)\,|\,t=t\,|\,\neg A \,|\,(A \wedge A)\,|\,(A \vee A)\,|\,(A \to A)\,|\,(\exists x)A \,|\,(\forall x)A.
$$
A formula of $\mathcal{L}$ that contains neither $\neg$ nor $\to$ is called \emph{positive}.
\end{definition}

\noindent
Constant symbols conceptually correspond to the ``initial functions'' of a function algebra, a mathematical object defined by a set of basic functions and schemes to construct new functions from given ones, commonly used in recursion theory to characterize complexity classes; see~\cite{clote:99}.
%
Related notions of open and closed terms are defined as is standard.
Moving to formulas, the predicate $\ms{W}$ intuitively means ``being a binary word'' and is needed to distinguish which terms represent elements of $\bb{W}$ and which do not.

A formal semantics associated with this language and consistent with the intuitive interpretation presented above has been detailed in~\cite[Sec.~$6.2$]{strahm:03} relying on the (open) term model $M(\lambda\eta^+)$ \cite[Sec.~$6.2$]{strahm:03}, which is based on the extended untyped lambda calculus with $\eta$ reduction (see~\cite{cantini:05}).
Full details can be found in~\cite{strahm:03,eberhard:15}.

\begin{notation}
    We use lower case letters to denote terms of $\mathcal{L}$, upper case letters to denote formulas or functions over (the Cartesian products of) $\bb{W}$, and Greek letters for elements of $\bb{W}$.
    To each word $\sigma \in \bb{W}$, we associate a canonical term $\overline{\sigma} \in \mathcal{L}$, constructed by application of $\ms{s_0}$ and $\ms{s_1}$ to $\epsilon$.
    We shorten term application $(t\cdot s)$ as the juxtaposition of terms $ts$, adopting the convention of association to the left (for $n\ge 2$ terms).
    We adopt standard conventions on formulas: left-precedence for iterations of $\wedge$ and $\vee$, which bind stronger than $\to$.

To enhance readability, we adopt the following abbreviations:
    $s\ms{0}:= \ms{s_0}s$ and $s\ms{1}:=\ms{s_1}s$ (when $s=\epsilon$, we simply write $\ms{0}$ and $\ms{1}$);
    $\ms{c_{\epsilon}}s\,t\,r:= \ms{c_W}s\,t\,r\,r$, which is $t$ if $s=\epsilon$ and $r$ otherwise (conditional on size);
     $s_1,\dots, s_n\in \ms{W}:= \ms{W}(s_1) \wedge \dots \wedge \ms{W}(s_n)$;
    $\forall x \in \ms{W}.A := (\forall x)(x\in \ms{W} \to A)$;
    $\exists x \in \ms{W}.A := (\exists x)(x\in \ms{W} \wedge A)$;
    %
    $f:\ms{W}^n\to \ms{W} := \forall x_1,\dots, x_n \in \ms{W}.\ms{W}(fx_1 \dots  x_n)$.
\end{notation}

The theories considered in this paper are all extensions of a slightly modified version of the basic theory introduced in~\cite{eberhard:15} (see App.~\ref{app:formulation_B}).
The only differences are the use of $\ms{c_W}$ (as~\cite{kahle:11}), instead of $\ms{c_\subseteq}$, and the direct application of terms as a total operation, instead of using the logic of partial terms~\cite{beeson:85}.
Both modifications are harmless. In particular, adding totality of applications only yields a stronger theory, for which one can still prove the desired upper bounds (see~\cite{jager:95,strahm:03}).
As an example of the non-logical axioms defining a constant, the conditional on words is governed by the following, which intuitively computes a case distinction on the first input:
\begin{align*}
    &\ms{c_W} \epsilon y z w = y\qquad\qquad\ms{W}(x)\to\ms{c_W} {(}\ms{s_0} x{)} y z w = z\, \wedge \,\ms{c_W} {(}\ms{s_1} x{)} y z w = w
\end{align*}

\begin{definition}\label{def:B}
    The theory $\ms{B}$ over 
    $\mathcal{L}$ 
    contains (i) all axioms of classical logic with equality and extensionality, and (ii) the universal closures of the non-logical axioms defining the constants of $\mathcal{L}$.
\end{definition}
\noindent
Observe that the theory $\ms{B}$ does not include any induction scheme.
The following basic properties of $\ms{B}$ will be used in the remainder of the paper.

\begin{lemma}[\cite{strahm:03}]\label{lemm:fixpoint}\,
\begin{itemize}
\itemsep0em
    \item \textit{Abstraction}: For each term $t\in\mathcal{L}$ and variable $x$, there is an $\mathcal{L}$-term $(\lambda x.t)$, whose free variables are those of $t$, excluding $x$, such that ${\sf B}\vdash (\lambda x.t) x = t$.
    \item \textit{Recursion}: There is a closed $\mathcal{L}$-term $\mathtt{rec}$, such that ${\sf B}\vdash \mathtt{rec}\, t = t\,(\mathtt{rec}\, t)$, for any term $t$.
\end{itemize}
\end{lemma}

\noindent
Note that $(\lambda x.t)$ simply denotes a term (constructed inductively~\cite[Prop.~$1$]{strahm:96}), rather than an object-level lambda abstraction.
However, the name is suggestive, as it precisely represents the term which, when applied to $s$, is provably equal to the term $t$ with all occurrences of $x$ replaced by $s$.
We often say that a term $t$ returns $s$ on input $x$ instead of writing $\mathsf{B} \vdash tx=s$. Similarly, we use $\lambda x_1 \dots x_n.t$ to denote the term $(\lambda x_1 \dots (\lambda x_n. t)\dots )$.

Lemma~\ref{lemm:fixpoint} implies that one can write $\mathcal{L}$-\emph{terms} representing any partial recursive function (see Prop.~\ref{prop:lower_bound_counting_TC} for an example). This is not the case
for theories developed in the context of Bounded Arithmetic~\cite{parikh, buss:85}, where the language is less expressive and functions are defined by \emph{formulas}.
More precisely, when dealing with applicative theories, we are not interested in the functions that can be \textit{represented} by some term (or formula), but rather in those for which the theory can \textit{prove} that the representing term is \textit{total}.

\begin{definition}\label{def:provably_total_function}
Let {\sf T} be an applicative theory extending {\sf B}.
A function $F:\bb{W}^n \to \bb{W}$ is \textit{provably total} in ${\sf T}$ if there exists a closed $\cal{L}$-term $f$ such that:
\begin{itemize}
\itemsep0em
    \item[(i)] for each $\sigma_1,\dots,\sigma_n\in\bb{W}$, ${\sf T} \vdash f\,\,\overline{\sigma_1}\dots\overline{\sigma_n} = \overline{F(\sigma_1,\dots,\sigma_n)}$
    \item[(ii)] ${\sf T}\vdash f:\ms{W}^n\to\ms{W}$.
\end{itemize}
The set of such functions is denoted by $\comp{T}$
(for `computational power'). 
If (i) is satisfied, we say that the term $f$ \textit{represents} the function $F$.\footnote{As the ``for each'' statement in the definition is not quantifying inside the theory, one could replace {\sf T} with {\sf B} in item (i). Accordingly, we 
do not have to specify the theory when saying that $f$ \textit{represents} $F$.}
If (ii) is satisfied, we say that the term $f$ is ${\sf T}$-\textit{provably total}, or \textit{total} if no confusion arises. 
\end{definition}
\section{Applicative theories, from small circuit classes to polynomial space}\label{sec:the_theories}
In this section, we introduce original applicative theories for multiple classes. In Sec.~\ref{sec:rev}, we present a simplified version of the theory for $\bff{FAC}^0$ originally introduced in~\cite{eberhard:15}.
This theory is especially relevant as it constitutes the basis for obtaining all our extensions \emph{in a uniform way}.
In Sec.~\ref{sec:appT}, we introduce the inductive schemes needed to obtain all other applicative theories. Finally, in Secs.~\ref{sec:lowerB} and~\ref{sec:upper_bounds}, we prove that the these theories actually characterize corresponding complexity classes involving counting and alternation.
\subsection{A theory for \texorpdfstring{$\bff{FAC}^0$}{FAC0}, revisited}\label{sec:rev}
As mentioned, recursion theory traditionally characterizes complexity classes by a set of initial functions and a set of operations under which they are closed~\cite{cobham:64,clote:99}. In the context of applicative theories, a similar distinction is made between basic elements, identified by constants of $\mathcal{L}$ that are axiomatized as total terms, and the operational machinery, obtained by endowing $\mathsf{B}$ with induction schemes corresponding to proofs of totality of varying complexity.
In other words, 
while initial functions of the algebra correspond to terms whose totality is given by the axioms of the corresponding applicative theory, recursion schemes are usually captured by induction axioms.

Notably, a natural candidate for characterizing classes of functions is full induction on quantifier-free positive formulas. However, this induction scheme is already too powerful to characterize complexity classes.
Indeed, when added to $\mathsf{B}$, the resulting theory is highly expressive, reaching the strength of Primitive Recursive Arithmetic \cite[Sec. 7.2]{strahm:03}. Intuitively, a primitive recursive equation defining $F$ can be captured by the induction scheme:
$$
\mathsf{W}(f(\eps, x)) \land \forall y.\Big(\mathsf{W}(f(y,x))\to \mathsf{W}(f(\mathsf{s_\ell}y,x))\Big) \to \forall y.\mathsf{W}(f(y,x))
$$
where $f$ is an $\mathcal{L}$-term representing $F$ (see \cite{strahm:03} for details).

To capture fine-grained complexity measures, we must restrict this scheme. Specifically, a natural constraint we consider is restricting induction to special classes of formulas.

\begin{definition}\label{def:easy}
Let $\mathcal{E}$ denote the class of formulas which are positive, do not contain the predicate $\ms{W}$ nor \textit{disjunctions}; let $\mathcal{E}^\vee$ be the extension of $\mathcal{E}$ including disjunctions on formulas.
\end{definition}

\noindent
The name $\mathcal{E}$ comes from \textit{easy}, as these formulas have trivial realizers (see Sec.~\ref{sec:realizability}).
The relevance of these specific classes will be fully clarified later (see, e.g., Remarks~\ref{rmrk:bounded_size_realiezr} and~\ref{rmrk:easy_realizer}).
Broadly speaking, disjunctions often make it possible to circumvent the restrictions imposed by the form of an induction scheme. In many classes, including the ones treated in this paper, formulas of the form $fx = \eps\,\vee\,...\,\vee\, fx={11}$ allow one to simulate alternating Turing machines.
By disallowing disjunctions, the computational content of the equations $fx = y$ in $\mathcal{E}$ formulas is determined solely by the complexity of the terms $f$, and one may safely interpret $y$ as the value being searched for.
Thus, restriction to $\mathcal{E}$ ensures that inductions behave ``as one would expect'', given the corresponding recursion principle.

To capture $\bff{FAC}^0$ we introduce the following, restricted inductive scheme.

\begin{definition}\label{def:conc_induction}
The scheme of \emph{concatenation induction on notation} $(\BSI)$ is defined as:
$$ \exists z\in\ms{W}.A(\ms{\epsilon},z)\wedge\forall y \in\ms{W}.\forall z.\Big(A(y,z)\to A(yi,z0)\vee A(yi,z1)\Big) \to \forall y \in\ms{W}.\exists z\in\ms{W}\,A(y,z)
$$
where $A(y,z)\in\cal{E}$ and $i\in\{0,1\}$ is used to shorten the same formula for $y0$ and $y1$.
\end{definition}

\noindent
Thinking of the formula $A(y,z)$ defined as $fyx=z$, the base of the induction says that there exists $z\in\ms{W}$ such that $f\epsilon x=z$ (the initial value) and the induction step that, for each $y\in\ms{W}$, if $fyx=z$, then $f(yi)x=z0$ or $f(yi)x=z1$ (concatenation). 
If one proves these two formulas, then the conclusion is the statement of the totality of $f$.
Thus, the scheme $(\BSI)$ has a clear analogy to Concatenation Recursion on Notation ({\sf CRN}) \cite{clote:99}, which is used to characterize the logarithmic-time hierarchy (equivalent to $\bff{FAC}^0$).
Recall that the recursion step in $\mathsf{CRN}$ is of the form $F(yi,x) = S_{H(yi,x)}(F(y,x))$, so $F(yi,x)$, defined from $H$, may concatenate $0$ or $1$ with its previous iteration, \emph{and} the decision of which bit to concatenate does \emph{not} depend on $F(y,x)$. It is solely determined by $H(yi,x)$. This is the reason why, in the scheme above, only $y$ is quantified over $\ms{W}$. This means that one proves the induction step for an arbitrary $y\in\ms{W}$, guaranteed to be a word as it is in $\ms{W}$, and for an arbitrary $z$, which could in theory be any $\mathcal{L}$-term.
Observe also that when one replaces $\forall z$ above with $\forall z\in\ms{W}$, then the resulting scheme rather captures Full Concatenation Recursion on Notation ($\mathsf{FCRN}$), where the recursion step is of the form $F(yi,x) = S_{H(yi,x,F(y,x))}(F(y,x))$. This can be used to characterize $\bff{FP}$~\cite{ishihara:99,eberhard:15}.
In Eberhard's characterization of $\bff{FAC}^0$, instead of $z$ ranging over the entire universe, a special predicate $\ms{V}$ was introduced and the above axiom was formulated with $\forall z\in\ms{V}$ (see \cite{eberhard:15} for details).

\begin{definition}\label{def:LHT}
The theory $\sf{LHT}$ is defined by endowing $\sf{B}$ with the axiom 
$(\BSI)$. 
\end{definition}

\noindent
The characterization proof for $\bff{FAC}^0$ proceeds smoothly by generalizing techniques from~\cite{eberhard:15}; see Sec.~\ref{sec:background} for a comparison between $\sf{LHT}$ and Eberhard's theory, and App.~\ref{app:alterations_AC0} and~\ref{sec:upper_bound_fac0} for proof details.

\begin{theorem}\label{thm:ac0_characterizatio}
$\comp{LHT} = \bff{FAC}^0$.
\end{theorem}

\begin{remark}\label{rmrk:quantifiers_not_W_ok}
Note that in the induction formula, arbitrary quantifiers are allowed, provided that they do not range over $\ms{W}$. The intuitive reason for this is that, by Def.~\ref{def:provably_total_function}, proving a function to be total requires the ability to introduce the predicate $\ms{W}$. By the axiomatization of $\ms{W}$, quantifiers do not help in doing this.
\end{remark}
\subsection{Theories with schemes of addition induction}\label{sec:appT}
To capture classes featuring counting or alternation, we introduce the following induction schemes, which intuitively allow us to either leave $z$ unchanged or add 1 to $z$.
The fact that we are considering induction by concatenation of either 0 or 1 (induction on notation), or by lexicographic successor ($\ms{s_\ell}$), changes the number of induction steps involved, clearly altering the corresponding expressive power.
Another crucial feature that changes the power of induction is the use of either easy formulas or formulas involving disjunction.
Notably, these apparently minimal syntactical changes to our `unique' induction scheme substantially modify its computational power (i.e.~the set of functions provably total in the theory).

\begin{definition}\label{def:counting_induction}\label{def:extended}
The scheme of addition induction $(\SSI)$ is defined as follows:
    $$\exists z\in\ms{W}.A(\epsilon,z)\wedge\forall y \in\ms{W}.\forall z.\Big(A(y,z)\to A(yi,z)\vee A(yi,\ms{s_\ell}z)\Big) 
    \to \forall y \in\ms{W}.\exists z\in\ms{W}\,A(y,z)$$
where $A(y,z)\in\cal{E}$ and $i\in\{0,1\}$ is used to abbreviate writing the same formula for $y0$ and $y1$.
We use $(\SSIl)$ to denote the lexicographic version of $(\SSI)$, defined below:
$$
    \exists z\in\ms{W}.A(\epsilon,z)\wedge\forall y \in\ms{W}.\forall z.\Big(A(y,z)\to A(\ms{s_\ell}y,z)\vee A(\ms{s_\ell}y,\ms{s_\ell}z)\Big) 
    \to \forall y \in\ms{W}.\exists z\in\ms{W}\,A(y,z) 
    $$
%
If $(Ind)$ stands for any of the above induction schemes, we use $(Ind)^\vee$ to denote the same scheme, where disjunctions are allowed on the induction formula.
\end{definition}

The theories considered in the following sections to characterize complexity classes are obtained by simply endowing the basic theory for $\bff{FAC}^0$ with one of the schemes described above.

\begin{definition}\label{def:theories}
The following theories are obtained by adding to {\sf LHT} the corresponding induction scheme:
\addtolength{\tabcolsep}{0.4em}
\begin{table}[H]
\centering
\label{tab:theories_def}
\begin{tabular}{c | c  c}
\toprule
 & \textbf{No disjunctions} & \textbf{Disjunctions} \\
\midrule
\stackanchor{\textbf{Induction}}{\textbf{on notation}} & $\sf{LCHT} = \sf{LHT} + (\SSI)$        & $\sf{ALT} = \sf{LHT} + (\SSI)^\vee$ \\[0.4cm]
\stackanchor{\textbf{Lexicographic}}{\textbf{induction}} & $\sf{CHT} = \sf{LHT}+ (\SSIl)$     &  $\sf{PST} = \sf{LHT} + (\SSIl)^\vee$\\
\bottomrule
\end{tabular}
\caption{Applicative theories}
\end{table}
\end{definition}

\vspace{-0.6cm}
\noindent
The names of the theories are suggestive of the 
classes they characterize (see Table \ref{tab:theories}). The next sections are devoted to proving the characterizations and showing that ${\sf ALT}\neq{\sf CHT}$.
\section{Lower bounds}\label{sec:lowerB}
One of the advantages of applicative theories is that lower bounds are straightforward to prove, starting from function algebraic characterizations for corresponding classes.
The key idea is to rely on Lemma~\ref{lemm:fixpoint} to define an $\mathcal{L}$-term representing a function constructed by a corresponding recursion principle, and then to use the available induction scheme to prove its totality (see~\cite{strahm:03}).
\subsection{Lower bounds for counting classes}\label{sec:lower_bound_tc0_ch}
For both $\mathsf{LCHT}$ and $\mathsf{CHT}$, the proof strategy relies on constructing a term that implements a form of counting.
Here, we crucially exploit the fact that $\bff{FTC}^0$ can be defined by simply closing $\bff{FAC}^0$ under bit-counting~\cite{chandra:84,buss:92,vollmer} and that $\bff{FCH}$ is obtained by closing $\bff{FAC}^0$ under summation of Boolean functions~\cite{vollmer:96,lago:22}.

\begin{proposition}\label{prop:lower_bound_counting_TC}
$\bff{FTC}^0\subseteq\comp{LCHT}$.
\end{proposition}
\begin{proof}
Since $\bff{FTC}^0$ is obtained by extending $\bff{FAC}^0$ with bit-counting \cite{chandra:84,buss:92}, we just have to show that in $\comp{LCHT}$ we can count bits. Define $\mathtt{bc} = \mathtt{rec}\,r$, where
$
    r = \lambda fy.\,\ms{c_W}y\epsilon(f\ms{p_w}y)(\ms{s_\ell}(f\ms{p_w}y)).
$
Clearly, the term $\mathtt{bc}$ represents the following recursive function:
\begin{align*}
    &BC(\eps) = \eps\qquad BC(y0) = BC(y)\qquad BC(y1) = BC(y)+ 1&
\end{align*}
where $+$ is used here with a slight abuse of notation, as it does not denote standard addition but the function corresponding to lexicographic addition.
%
Indeed, recalling that $\ms{c_W}$ is axiomatized by $\ms{c_W} \epsilon y z w = y$ and $\ms{W}(x)\to\ms{c_W} {(}\ms{s_0} x{)} y z w = z\,\,\wedge\,\,\ms{c_W} {(}\ms{s_1} x{)} y z w = w$, the following identities are easily provable in {\sf B}, under the assumption that $\ms{W}(y)$ holds:
\begin{align*}
    &\mathtt{bc}\,\epsilon = \mathtt{rec} \, r\epsilon  = r(\mathtt{rec} \, r)\epsilon = r\mathtt{bc}\,\epsilon  = \ms{c_W}\epsilon\epsilon(\mathtt{bc}\,\ms{p_w}\epsilon)(\ms{s_\ell}(\mathtt{bc}\,\ms{p_w}\epsilon)) = \epsilon&\\
    &\mathtt{bc}(y0) =  r\mathtt{bc}(y0) = \ms{c_W}(y0)\epsilon(\mathtt{bc}\,\ms{p_w}(y0))(\ms{s_\ell}(\mathtt{bc}\,\ms{p_w}(y0))) = \mathtt{bc}\,(\ms{p_w}(y0)) = \mathtt{bc}\,y&\\
    &\mathtt{bc}(y1) = r\mathtt{bc}(y1) = 
     \ms{c_W}(y1)\epsilon(\mathtt{bc}\,\ms{p_w}(y1))(\ms{s_\ell}(\mathtt{bc}\,\ms{p_w}(y1))) =
    \ms{s_\ell}(\mathtt{bc}\,y).&
\end{align*}
So, $\mathtt{bc}\,y$ returns the number of $1$'s in $y$, written in $\bb{W}$ notation.
Note that, to deal with $\mathtt{bc}(y0)$ and $\mathtt{bc}(y1)$, we use the fact that $y$ is quantified over $\ms{W}$, as this is required to apply the second axiom of $\ms{c_W}$.
Then, to show it provably total in {\sf LCHT}, we consider the induction formula $A(y,z):= \mathtt{bc}\,y=z$
For $y=\epsilon$, $\exists z\in\ms{W}.A(\epsilon,z)$ holds, since, for $z = \epsilon$, $\mathtt{bc}\,\epsilon = z$.
Now let $y\in\ms{W}$ and assume that $A(y,z)$ holds, i.e., that $\mathtt{bc}\,y = z$.
Then, by construction, $\mathtt{bc}(y0) = \mathtt{bc}\,y = z$ and $\mathtt{bc}(y1) = \ms{s_\ell}(\mathtt{bc}\,y) = \ms{s_\ell}z$, so we have $A(yi,z)\vee A(yi,\ms{s_\ell}z)$.
Thus, from the induction scheme $(\SSI)$, the totality of $\mathtt{bc}$ follows.
\end{proof}

\begin{proposition}\label{prop:lower_bound_counting}
$\bff{FCH}\subseteq\comp{CHT}$.
\end{proposition}
\begin{proof}
Analogous to the proof of Prop.~\ref{prop:lower_bound_counting_TC}, but relying on the fact that $\bff{FCH}$ can be obtained as the closure of $\bff{FAC}^0$ under $Sum [H](y,x) = \sum_{z\leq y}H(z,x)$~\cite[Th.~4.3]{vollmer:96}, even when $H$ is assumed to return only $\eps$ or $0$ ($0$ and $1$ in binary notation, respectively) (since adding such functions suffices to compute the operator of Def.~\ref{def:classes}); see, e.g.~\cite[Lemma 3.2(3)]{wagner:86}. 
In particular, for $h:\ms{W}^2\to\{\eps,0\}$ representing a function $H$, consider the term $\mathtt{sum} = \mathtt{rec}\,r$, where $r = \lambda fyx.\,\ms{c_\epsilon}y(h\eps x)(\ms{c_W}(hyx)(f(\ms{p_\ell}y)x) (\ms{s_\ell}(f(\ms{p_\ell}y)x))(\ms{s_\ell}(f(\ms{p_\ell}y)x)))$, which represents the recursive function $SUM$, s.t.~$SUM(\eps,x) = H(\eps,x)$ and $SUM(y+ 1,x) = H(y+ 1,x) + SUM(y,x)$.
With the induction formula $A(y,x,z):= \mathtt{sum}\, yx=z$, the totality of $\mathtt{sum}$ can be proven from $(\SSIl)$.
Thus, since $\comp{CHT}$ contains $\bff{FAC}^0$ and is closed under $Sum$, it contains $\bff{FCH}$ as well.
\end{proof}
\subsection{Lower bounds for alternating classes}\label{sec:lower_bound_nc1_psapce}
Lower bounds for theories corresponding to the classes $\mathbf{FNC}^1$ and $\mathbf{FPSPACE}$ are proved following similar ideas. The main difference is that here disjunctions are allowed in the induction formula, making induction expressive enough to simulate a form of constant-bounded recursion; 
that is, a recursive definition where $|F(y,x)|$ is required to be bounded by a constant $k$.
This, in turn, allows us to characterize the computation performed by alternating classes.
Indeed, recall that the class $\mathbf{FPSPACE}$ (resp., $\mathbf{FNC}^1$) has been characterized in terms of function algebras crucially including a scheme called $k$-bounded recursion, $k$-$\mathsf{BR}$ (resp., on notation, $k$-$\mathsf{BRN}$), for $k \ge 4$:
a function $F$ is defined by $k$-$\mathsf{BR(N)}$ from $G$ and $H$ if
$F(\eps, x)=G( x)$ and $F(y+1, x)=H(y, x,F(y,x))$
(resp., $F(yi,x)=H_i(y,x,F(y,x))$, for $i\in \{0,1\}$),
where $y+1$ denotes $\ms{s_\ell}y$, and $F$ takes only values in $\{0,\dots, k\}$
~\cite{clote:93,clote:97} (see App.~\ref{sec:algebras} for further details).

\begin{proposition}\label{prop:lower_bound_pspace}
$\bff{FPSPACE}\subseteq \comp{PST}$.
\end{proposition}
\begin{proof}
The proof consists in showing that $\comp{PST}$ is closed under constant-bounded recursion, $k$-$\mathsf{BR}$.
The result is obtained for the bound $k=4$, i.e.,~for size bounded by $2$ (recall that in $\bb{W}$, $4$ is $01$, see Notation~\ref{not:notation}), as this is already enough to capture $\bff{FPSPACE}$~\cite{clote:97,clotetakeuti}.
%
Let $G, H\in\comp{PST}$, and $F$ be defined by $k$-$\mathsf{BR}$ from $G,H$, with bound $k=4$.
Let $g$ and $h$ be ${\sf PST}$-provably total functions representing $G$ and $H$ (in the sense of Def.~\ref{def:provably_total_function}).
First, we consider the auxiliary $\ms{b_2}x := \ms{bit}0x\,\ms{*}\ms{bit}\eps x$, which returns the two right-most bits of $x$.\footnote{Note that, if $|x|=1$, $\ms{bit}0x = \eps$, so only the right-most bit of $x$ remains, and if $x=\eps$, then $\ms{b_2}x=\eps$ as well. That is, the function $\ms{b_2}x$ needs not always return two bits.}
Then, we define the term $f = \mathtt{rec}\,r$, where
    %
    $r = \lambda fyx.\,\ms{c_\epsilon}\,y\,\,(\ms{b_2}(gx))\,(\ms{b_2}(h(\ms{p_\ell}y)x(f(\ms{p_\ell}y)x)))
    $
and representing the function:
\begin{align*}
    &F'(\eps,x) = BIT(0,G(x)) * BIT(\eps,G(x))\\
    &F'(y+1,x) = BIT(0,H(y,x,F(y,x))) * BIT (\eps,H(y,x,F(y,x)))&
\end{align*}
where $*$ is the concatenation function and $BIT(z,w)$ returns the $z$th bit of $w$, or $\eps$, if $z\geq|w|$ or $w = \eps$.
Clearly, the function $F'$ always returns only the last two bits of the computation performed by $G$ and $H$.
Since $|F(y,x)|\leq 2$, $F'(y,x)=F(y,x)$ for all $y,x$.
Now let $A(y,x,z)$ be the formula $z=z\wedge (fyx = \epsilon\vee fyx = 0\vee \cdots\vee fyx=11)$, which states that $fyx$ has a value of size at most $2$.
We use $(\SSIl)^\vee$ to prove the totality of $f$.\footnote{Note that this formula is not allowed in the induction scheme $(\SSIl)$.}
Fix the parameter $x\in\ms{W}$. 
For $y=\epsilon$, $\exists z\in\ms{W}.A(\epsilon,x,z)$ holds since $f\epsilon x = \ms{b_2}(gx)$ and $g$ is provably total by assumption.
Let $y\in\ms{W}$ and $z$ be arbitrary and suppose that $A(y,x,z)$ holds.
Then, $fyx =\epsilon\vee\cdots\vee fyx=11$, so, in particular, $fyx\in\ms{W}$.
From this, together with the totality of $h$, it follows that $hyx(fyx)\in\ms{W}$, so $f(\ms{s_\ell}y)x = \ms{b_2}(hyx(fyx))$ also has value between $\epsilon$ and $11$.
Thus, $z=z\wedge (f(\ms{s_\ell}y)x = \epsilon\vee f(\ms{s_\ell}y)x = 0\vee \cdots\vee f(\ms{s_\ell}y)x=11)$ holds, i.e., $A(\ms{s_\ell}y,x,z)$ holds.
Thus, from $(\SSIl)^\vee$, the totality of $f$ follows.
\end{proof}

In a completely analogous fashion, one can show that $\comp{ALT}$ is closed under $k$-$\mathsf{BRN}$~\cite{clote:93} (see \cite[Lemma 6]{eberhard:15}), obtaining the following result:

\begin{proposition}\label{prop:lower_bound_nc1}
$\bff{FNC}^1\subseteq \comp{ALT}$.
\end{proposition}
\section{Upper bounds}\label{sec:upper_bounds}
In this section, we complete the characterization proofs by considering upper bounds.
Usually, to prove that $\comp{T} \subseteq \mathbf{FC}$, for some applicative theory $\mathsf{T}$ capturing a complexity class of functions $\mathbf{FC}$, one proceeds as follows (see e.g.~\cite{strahm:03,eberhard:15}):
\begin{itemize}
\itemsep0em
    \item[(i)] formulate {\sf T} 
    as a sequent calculus and prove that if ${\sf T}\vdash\Gamma\To\Delta$, $\Gamma, \Delta$ containing only positive formulas, there is a {\sf T}-derivation of $\Gamma\To\Delta$ where only positive formulas appear;
    \item[(ii)] prove that each axiom of the sequent formulation of {\sf T} has a corresponding realizing function in $\bff{FC}$, and that, whenever a rule is applied to a set of premises containing only positive formulas, which are realized by functions in $\bff{FC}$, then the conclusion has a realizing function in $\bff{FC}$ as well.
\end{itemize}
\noindent
Since the statement of totality of a function will have the form $\ms{W}(x)\To\ms{W}(fx)$, item (i) implies that every function whose totality is provable in {\sf T} has a {\sf T}-derivation where only positive formulas appear. Positivity ensures that the construction of realizers is simple (for more complicated cases in which non-positive formulas occur, see \cite{strahm:04}).
Item (ii) gives an inductive construction of a realizing function for the sequent $\ms{W}(x)\To\ms{W}(fx)$, starting from realizers for the axioms, and constructing new realizers for each step of the proof of $\ms{W}(x)\To\ms{W}(fx)$.
See also Remark \ref{rmrk:why_realizability_makes_sense}.
\subsection{Sequent calculus formulation}\label{sec:cut_elimination}
Sequent calculi for our theories are defined in the standard way in all their logical parts (see e.g.~\cite{gentzen:35,troelstra:00}).
The original aspect lies in the non-logical axioms and induction rules, which correspond to the induction schemes defined in the lemma below.

\begin{lemma}\label{lemm:sequent_calculus}
For ${\sf T}\in\{{\sf LCHT},{\sf ALT},{\sf CHT},{\sf PST}\}$, there is a sequent calculus formulation of ${\sf T}$ where all the main formulas of non-logical axioms and rules are positive.
\end{lemma}
\begin{proof}
The idea is to convert each universal proposition into all substitution instances, each implication $A\to B$ into the sequent $A\To B$ and each negation $\neg A$ into the sequent $A\To$.
The treatment for axioms is standard and has been extensively considered (see e.g.~\cite[Sec. 6.1]{strahm:03}).
The formulation for the induction rule corresponding to $(\BSI)$ resembles the formulation given in~\cite{eberhard:15} for the corresponding scheme of induction:\footnote{As usual in sequent formulations, we take all substitution instances of an axiom. This ensures, for example, that we can use cut on, say, $\ms{W}(g(s))$, to deduce $\ms{W}(s)\To \ms{W}(h(g(s)))$ from $\ms{W}(s)\To \ms{W}(g(s))$ and $\ms{W}(g(s))\To \ms{W}(h(g(s)))$.
}
\begin{prooftree}
\AxiomC{$\Gamma\,\,\To\,\,\exists z\in\ms{W}. A(\epsilon,z),\Delta$}
\AxiomC{$\Gamma,\ms{W}(y),A(y,z) \,\,\To\,\, A(yi,z0),A(yi,z1),\Delta$}
 \RightLabel{$R_{(\BSI)}$}
    \BinaryInfC{$\Gamma,\ms{W}(s)\,\, \To\,\,\exists z\in\ms{W}\,A(s,z),\Delta$}
\end{prooftree}
where $i$ ranges over $\{0,1\}$ (so the rule has three premises, see App.~\ref{sec:upper_bounds_appendix}), $y$ and $z$ denote fresh variables, not occurring in $\Gamma$ or $\Delta$, and $A(y,z)\in\cal{E}$.

Moving to the summing induction schemes, we define them in a similar way.
We consider here only the rule corresponding to $(\SSIl)^\vee$ as an example:
\begin{prooftree}
    \AxiomC{$\Gamma\,\,\To\,\,\exists z\in\ms{W}. A(\epsilon,z),\Delta$}
    \AxiomC{$\Gamma,\ms{W}(y),A(y,z) \,\,\To\,\, A(\ms{s_\ell}y,z),A(\ms{s_\ell}y,\ms{s_\ell}z),\Delta$}
    \RightLabel{$R_{(\SSIl)^\vee}$}
    \BinaryInfC{$\Gamma,\ms{W}(s)\,\, \To\,\,\exists z\in\ms{W}\,A(s,z),\Delta$}
\end{prooftree}
where $y$ and $z$ denote fresh variables, not occurring in $\Gamma$ or $\Delta$, and $A(y,z)\in\cal{E}^\vee$.
Since formulas in $\cal{E}^\vee$ and $\cal{E}$ are, in particular, positive, we obtain a formulation of the induction schemes where all main formulas are positive.  
\end{proof}
It is already clear where the condition that the variable $z$ should be quantified over the entire universe comes in. The ``value'' of $z$, i.e.,~$\ms{W}(z)$, is not given in the left-hand side of the second premise. When we move to realizability, this will mean that the realizing function is able to realize the right-hand side without a realizer for $\ms{W}(z)$.

By a generalization of Gentzen's cut-elimination for {\sf LK}~\cite{gentzen:35}, the cut rule can be eliminated in these theories up to positive formulas: cuts on non-positive formulas, as well as logical rules, are reduced in the standard way. 
We thus obtain the following result.

\begin{theorem}\label{thm:positive_suffices_T}
For ${\sf T}\in\{{\sf LCHT},{\sf ALT},{\sf CHT},{\sf PST}\}$, if ${\sf T}\vdash \Gamma\To\Delta$, for $\Gamma,\Delta$ containing only positive formulas, then there is a {\sf T} derivation of $\Gamma\To\Delta$ containing only positive formulas.
\end{theorem}

Since the statement of totality of a function is of the form $\ms{W}(x)\To\ms{W}(fx)$, it follows that proofs where only positive formulas appear already capture the \textit{computational power} of the corresponding theory: if $\sf{T}$ proves the totality of a function, then there exists a $\sf{T}$ derivation of totality where only positive formulas appear.
Intuitively, in these proofs, non-logical axioms will correspond to axiomatizing initial functions as total, induction schemes will correspond to recursion schemes, and the cut rule \emph{on positive formulas} will correspond to function composition (Prop.~\ref{prop:oracle_in_theory}).
\subsection{Realizability}\label{sec:realizability}
The relevant property of proofs containing only positive formulas is that it is straightforward to extract from them a program as a realizing function. 

\begin{definition}\label{def:realizability}
For a positive formula $A$ and a word $\rho\in\bb{W}$, we write $\rho \rhdd A$ to express that $\rho$ ``realizes'' $A$ in the sense of \cite[Sec. $6.2$]{strahm:03} (see Def. \ref{def:realizability_appendix}).
\end{definition}

\noindent
This notion is generalized to \textit{sequents} through \textit{realizing functions}.

\begin{definition}[\cite{strahm:03}]\label{def:realizing_function}
Let $\Gamma = \{A_1,\dots,A_n\}$ and $\Delta$ be sets of positive formulas, whose free variables are in the list $\vec{u}$. The sequent $\Gamma\To\Delta$ is \textit{realized} by a function $F$ if, for all terms $\vec{s}$ and all words $\rho_1,\dots,\rho_n \in\bb{W}$, whenever for every $i \leq n$, $\rho_i \rhdd A_i[\vec{s}\,]$, it holds that $F(\rho_1,\dots,\rho_n) \rhdd \Delta[\vec{s}\,]$.
The function $F$ is called the \textit{realizing function} of $\Gamma\To\Delta$. 
\end{definition}
\noindent
We interpret a sequent $\Delta$ disjunctively, so $\rho\rhdd\Delta$ if $\rho=\pair{i,\rho'}$, where $i$ is an index of a formula in $\Delta$, which is realized by $\rho'$.

\begin{remark}\label{rmrk:why_realizability_makes_sense}
Suppose that $f$ is an $\cal{L}$-term representing $F:\bb{W}\to\bb{W}$ and that the sequent $\ms{W}(x)\To\ms{W}(fx)$ is realized by $G:\bb{W}\to\bb{W}$.
Then, for any $\rho\in\bb{W}$, $\rho\rhdd\ms{W}(\overline{\rho})$, whence $G(\rho)=\pair{0,\rho'}$ and $\rho'\rhdd\ms{W}(f\,\overline{\rho})$. 
As $f$ represents $F$, in the sense of Def.~\ref{def:provably_total_function}, it follows that, for every $\rho\in\bb{W}$, $(G(\rho))_1=F(\rho)$.
Thus, the complexity of the \textit{realizing function} $G$ upper bounds the complexity of $F$.
\end{remark}

Moving to the realizability of the theories from Sec.~\ref{sec:the_theories}, we start by considering the smallest theory $\sf{LHT}$, for which Th.~\ref{thm:realizability_known} is established following the lines of the proof in~\cite{eberhard:15}.

\begin{theorem}\label{thm:realizability_known}
Let $\Gamma\To\Delta$ be a sequent where only positive formulas appear. Then, if ${\sf LHT}\vdash \Gamma\To\Delta$, there is an $\bff{FAC}^0$ realizing function for $\Gamma\To\Delta$.
\end{theorem}

\noindent
In the next section, we extend this result to the remaining theories. Following~\cite{kahle:11,eberhard:15}, these proofs exploit the following two important properties of $\cal{E}$ and $\cal{E}^\vee$ formulas:

\begin{remark}\label{rmrk:easy_realizer}
If $A$ is a true $\mathcal{E}$-formula, then $A$ is realized by a fixed word $\rho_A$ that
depends only on the structure of $A$, not on its terms, and consists solely of pairings
of the empty word. For example, if $tx = y\,\wedge\,fz = l$ holds, then $\pair{\eps,\eps}$
realizes it, independently of the terms $tx$, $y$, $fz$ and $l$. In general,
Def. \ref{def:realizability} gives the construction of these constant realizers.
\end{remark}

\begin{remark}\label{rmrk:bounded_size_realiezr}
If $A$ is an $\cal{E}^\vee$ formula, the size of the realizer for $A$ (i.e., the number of bits of the word realizing $A$) is bounded by a constant that is independent of the terms in $A$.
For example, a realizer for $tx = y\,\vee\,fz = l$ is either $\pair{0,\eps}$ or $\pair{1,\eps}$, independently of these terms.
\end{remark}
\subsection{Proof of upper bounds}\label{sec:upper_bounds_proofs}
For $\bff{FTC}^0$ and $\bff{FCH}$, our proofs rely on the fact that the variable $z$, intuitively representing the output of a function, is not ranging over $\ms{W}$.
This ensures that we are able to distinguish \emph{in the desired complexity class} whether a side formula is realized (instead of the main one), and in this case return a realizer for it. 
Due to the restrictive nature of the schemes defining $\bff{FTC}^0$ and $\bff{FNC}^1$, this procedure has to be ``pre-processed'' by an auxiliary function, described below for the case of the small circuit classes (see~\cite[Lemma 11]{eberhard:15} for a similar remark).

\begin{definition}
Consider an instance of $R_{(\SSI)}$ (the case of $R_{(\SSIl)}$ is analogous)
such that $G,H_0$ and $H_1$ are realizing functions for the respective premises. 
Since $A\in\cal{E}$, let $\rho_A$ be a realizer for it (Remark~\ref{rmrk:easy_realizer}).
We define the following algorithm:\footnote{The index in $H_{\sigma(|\sigma|-|\tau|-1)}$ is made to ensure that we call the correct realizing function. For example, if $\sigma = 101$ and $\tau = 1$, then the correct realizing function to call is $H_0$ (as $\tau$ will be expanded with $0$ to still be a prefix of $\sigma$), which is obtained by $\sigma(|\sigma|-|\tau|-1) = \sigma(3-1-1) = \sigma(1) = 0$.}

\begin{algorithm}[H]
\SetArgSty{upshape}
\DontPrintSemicolon
\caption{Realizer for the side formulas}\label{alg:side_formulas}
\small
\textbf{Input} $\vec{\rho},\sigma$\;

\If{$(G(\vec{\rho}))_0\neq 0$}{
    \Return $G(\vec{\rho})$\;
}

\If{there is a proper prefix $\tau$ of $\sigma$ such that $(H_{\sigma(|\sigma|-|\tau|-1)}(\vec{\rho},\tau,\rho_A))_0\notin\{0,1\}$}{
    Let $\tau$ be the shortest such prefix and $j:=(H_{\sigma(|\sigma|-|\tau|-1)}(\vec{\rho},\tau,\rho_A))_0$\;
    \Return $\pair{j-1,\,(H_{\sigma(|\sigma|-|\tau|-1)}(\vec{\rho},\tau,\rho_A))_1}$\;
}
\Return $\epsilon$
\end{algorithm}
\end{definition}

\normalsize

\begin{lemma}\label{lemm:side_formulas}
Given an application of any of the rule $R_{(\SSI)}$, $R_{(\BSI)}$ or $R_{(\SSI)^\vee}$, Algorithm~\ref{alg:side_formulas} (resp.,~ with $\rho_A$ replaced by $Q(\vec{\rho},\tau)$ in the disjunctive cases, see~Lemma~\ref{lemm:fnc1_if_main_formula}) returns $\epsilon$ or a realizer for a side formula in $\Delta$.
%
\end{lemma}
\begin{proof}
Let $F'$ be the function computed by Algorithm \ref{alg:side_formulas}.
Suppose that $\vec{\rho}$ and $\sigma$ are realizers for the formulas in $\Gamma[\vec{s}]$ and $\ms{W}(s)$. We need to show that $F'(\vec{\rho},\sigma) = \eps$ or $F'(\vec{\rho},\sigma)\rhdd\Delta[\vec{s}]$.
If $(G(\vec{\rho}))_0\neq 0$, then $G(\vec{\rho})_0 = j$ is a pointer for a formula in $\Delta[\vec{s}]$ which is realized by $G(\vec{\rho})$. Thus, $F'(\vec{\rho},\sigma)\rhdd \exists z\in\ms{W}.\,A(s,z),\Delta[\vec{s}]$, as it returns the same pointer and realizer.
Otherwise, let $\tau$ be the smallest prefix of $\sigma$ for which $(H_{\sigma(|\sigma|-|\tau|-1)}(\vec{\rho},\tau,\rho_A))_0\notin \{0,1\}$, i.e., for which $H_{\sigma(|\sigma|-|\tau|-1)}(\vec{\rho},\tau,\rho_A)$ returns a pointer and a realizer for a formula in $\Delta[\vec{s}]$.
Since $\tau$ is the smallest such value, this means that, for all predecessors of $\tau$, the main formulas were always realized.
Since we know that $\exists z\in\ms{W}.A(\epsilon,z)$ is true, the above implies that, always, for each $\nu \subseteq \tau$, there is some term $r$ for which $A(\overline{\nu} ,r)$ is \textit{true}.
For the case of $\tau$ itself, let $r$ be this term.
Then the realizing function $H_{\sigma(|\sigma|-|\tau|-1)}(\vec{\rho},\tau,\rho_A)$ has the inputs:
\begin{enumerate}
\itemsep0em
    \item $\vec{\rho}$ realizing the formulas of $\Gamma[\vec{s}]$
    \item $\tau$ realizing $\ms{W}(\overline{\tau})$ (Def. \ref{def:realizability})
    \item $\rho_A$ realizing $A(\overline{\tau},r)$, since $A(\overline{\tau},r)$ is true and its realizer is a fixed word.
\end{enumerate}
Since the variable $z$ is not quantified over $\ms{W}$ and does not occur in $\Gamma$ or $\Delta$, this means that $H_{\sigma(|\sigma|-|\tau|-1)}(\vec{\rho},\tau,\rho_A)$ returns a correct realizer for $\Delta[\vec{s},z/r] = \Delta[\vec{s}]$.
Thus, $F'(\vec{\rho},\sigma)$ returns a realizer for $\Delta[\vec{s}]$.
Finally, if none of the above conditions are met, then it must be the case that the main formulas are always realized and $F'(\vec{\rho},\sigma) =\eps$.
\end{proof}

\noindent
Note that we critically used two aspects of the induction scheme $(\SSI)$: 
that $z$ does not range over $\ms{W}$, so that the realizing function does not need its value,
and that $A\in\mathcal{E}$, so that we can trivially produce a realizer for it.
The fact that, when $z$ ranges over $\ms{W}$, one obtains $\bff{FP}$ (\cite{ishihara:99,eberhard:15}), and that when disjunctions are allowed in $A$, one obtains $\bff{FNC}^1$ (\cite{eberhard:15} and below) shows that these conditions are indeed necessary, unless a collapse between the complexity classes is proven.

To complete the proof of the theories' upper bounds we will use the fact that the classes considered are closed under (sharply) bounded minimization, so that the above pre-processing can be implemented in the class (for $\bff{FTC}^0$ and $\bff{FNC}^1$, see e.g.~\cite[Def.~3.9]{clote:99}). Then, assuming that the side formulas are never realized, a realizer for the main formula will be easy to construct.
\subsubsection{Upper bounds for counting classes}\label{sec:realCircuits}
\begin{proposition}\label{prop:upper_bound_tc0}
If the premises in an application of $(\SSI)$ are realized by $\bff{FTC}^0$ functions, so is the conclusion.
\end{proposition}
\begin{proof}
Consider an instance of $R_{(\SSI)}$ and suppose that the premises are realized by $G,H_0,H_1\in\bff{FTC}^0$.
Since, whenever the main formula is not realized, we can construct a realizer for side formulas in $\bff{FTC}^0$ (Lemma \ref{lemm:side_formulas}), which is closed under sharply bounded minimization, we focus on the case where the main formula is realized. 
Assume that the main formula is realized in an instance of $R_{(\SSI)}$
with realizing functions $G$, $H_0$ and $H_1$ for the three premises.

Then, in order to construct our realizer we first define by {\sf CRN} (under which $\bff{FTC}^0$ is closed, see~\cite{clote:99}) the following function: $Q(\vec{\rho},\ms{\epsilon}) = \epsilon$ and $Q(\vec{\rho},\sigma i) =  S_{H_i(\vec{\rho},\sigma,\rho_A)_0}(Q(\vec{\rho},\sigma))$,
where $S_k$, for $k\in \{0,1\}$, is the binary successor over words (see App.~\ref{app:preliminaries}).
Intuitively, at each step, $H_i(\vec{\rho},\sigma,\rho_A)_0\in\{0,1\}$ will indicate whether the value of $z$, in the induction step, was incremented or not. 

Thus $\mathtt{BITSUM}(Q(\vec{\rho},\sigma))$, i.e., the number of $1$'s in
$Q(\vec{\rho},\sigma)$ (computable by constant-depth threshold circuits,
see~\cite{vollmer}), is precisely the number of times that the value of $z$ was incremented.
Thus, in $\bff{FTC}^0$ we can compute both: (1) the initial value $((G(\vec{\rho}))_1)_0$ of $z$, and (2) the number of times that this value is incremented.
Adding these values together, we get the realizer for $\ms{W}(z)$ in the conclusion.
If this value is $\tau$, we have that $\pair{0,\pair{\tau,\rho_A}}\rhdd \exists z\in\ms{W}.\,A(s,z),\Delta[\vec{s}]$.
\end{proof}

To prove the upper bound for {\sf CHT}, one can use a similar argument but considering counting machines in $\bff{FCH}$ (see Prop \ref{prop:upper_bound_CH}).
In particular, instead of concatenating all values of $H_i$, one considers a non-deterministic Turing machine which guesses a value $\tau$ smaller than $\sigma$ and accepts iff $(H(\vec{\rho},\tau,\rho_A))_0 = 1$.
The number of accepting configurations of this machine will be the number of times that the initial value is incremented, which can be calculated in $\bff{FCH}$. See App.~\ref{app:uppBCounting} (and Prop.~\ref{prop:upper_bound_CH}) for the proof.

\begin{proposition}\label{prop:upper_bound_fch}
If the premises in an application of $(\SSIl)$ are realized by $\bff{FCH}$ functions, so is the conclusion.
\end{proposition}
\subsubsection{Upper bounds for alternating classes}
The argument to prove upper bounds for $\bff{FNC}^1$ and $\bff{FPSPACE}$ is slightly different from the one from Sec.~\ref{sec:realCircuits}, as the induction formula $A$ no longer has a trivial realizer. 
Instead, we will use Remark~\ref{rmrk:bounded_size_realiezr}, establishing that realizers for $A$ have a bounded size.
Intuitively, the idea is 
to combine the previous strategy with the fact that, whenever the main formula is realized, its realizer can be computed in $\bff{FNC}^1$. This is essentially the strategy employed in \cite{eberhard:15}, which will yield a different algorithm for computing the side formulas.

\begin{lemma}\label{lemm:fnc1_if_main_formula}
If the premises in an application of $R_{(\SSI)^\vee}$ are realized by $\bff{FNC}^1$ functions, then there is a $\bff{FNC}^1$ function which returns a realizer for the main formula of the conclusion, whenever the main formulas of the premises are realized.
\end{lemma}
\begin{proof}
Since $A\in\cal{E}^\vee$, realizers for $A$ will depend on its terms, but their size will always be bounded by a constant $c_A$ (Remark~\ref{rmrk:bounded_size_realiezr}). 
We construct two functions, one which returns realizers for $A$, defined by $k$-{\sf BRN}, and another which will construct the variable $z$, defined by {\sf CRN}.
Define the following function
\begin{align*}
    &Q(\vec{\rho},\epsilon) = ((G(\vec{\rho}))_1)_1\qquad Q(\vec{\rho},\sigma i) = (H_i(\vec{\rho},\sigma, Q(\vec{\rho},\sigma)))_1
\end{align*}
which is in $\bff{FNC}^1$ (see~\cite{clote:93} or App.~\ref{app:preliminaries}).
At each recursion step, if $H_i(\vec{\rho},\sigma, Q(\vec{\rho},\sigma))$ returns a realizer for the main formulas, then its output, the realizer, is bounded in size by a constant. Analogously, so is $((G(\vec{\rho}))_1)_1$, whence the function $Q$ can be defined by $k$-{\sf BRN} for a suitable constant.

Analogously to the proof of Prop. \ref{prop:upper_bound_tc0}, we can now define a function $I(\vec{\rho},\sigma)$ via {\sf CRN} and $\mathtt{BITSUM}$ (both doable in $\bff{FNC}^1$), which returns a realizer for $\ms{W}(z)$ in the formula $\exists z\in\ms{W}.A(s,z)$.
The only difference is that now $I$ will use $Q(\vec{\rho},\sigma)$, defined above, to construct the realizers for $A$, instead of simply using $\rho_A$.
Therefore, the following realizing function is in $\bff{FNC}^1$: $F(\vec{\rho},\epsilon) = G(\vec{\rho})\,\,;\,\,F(\vec{\rho},\sigma i) = \pair{0,\pair{I(\vec{\rho},\sigma i),Q(\vec{\rho},\sigma i)}}$.
\end{proof}

\noindent
We conclude by proving the upper bounds for $\sf{ALT}$ and $\sf{PST}$.

\begin{proposition}\label{prop:upper_bound_nc1}\label{prop:upper_bound_pspace}
If the premises in an application of $R_{(\SSI)^\vee}$ (resp.,~$R_{(\SSIl)^\vee}$) are realized by $\bff{FNC}^1$ (resp.,~$\bff{FPSPACE}$) functions, so is the conclusion.
\end{proposition}
\begin{proof}
Consider an instance of $R_{(\SSI)^\vee}$ and suppose that the premises are realized by $\bff{FNC}^1$ functions $G,H_0,H_1$.
We start by running an adapted version of Algorithm~\ref{alg:side_formulas}, where the output $Q$ substitutes the constant realizer $\rho_A$.
If the first prefix $\tau$ is found, for which the side formulas are realized, then, up until $\tau$, it will always be the main formulas which are realized, and so $Q(\vec{\rho},\tau)$ returns a correct realizer for $A$, and so $H_i$ will return a correct realizer for $\Delta[\vec{s}]$.
Thus, if Algorithm~\ref{alg:side_formulas} does not return $\epsilon$, we have a realizer for the side formulas of the conclusion;
otherwise, if Algorithm~\ref{alg:side_formulas} does return $\eps$, then we know that the main formulas in the premises are always realized and thus we can run the algorithm described above to realize the main formula in the conclusion. 
This requires only a case distinction on two $\bff{FNC}^1$ algorithms, whence the final realizing function is in $\bff{FNC}^1$ as well.


For the theory {\sf PST}, the proof is similar to the upper bound for the $\bff{CH}$ (Prop. \ref{prop:upper_bound_CH}) and we do not need to run Algorithm \ref{alg:side_formulas}.\footnote{Alternatively, notice that Strahm's theory characterizing $\bff{FPSPACE}$ in~\cite{strahm:03} is stronger than ours.}
\end{proof}
\section{Separating theories via higher-order totality statements}\label{sec:separatin_oracle}
In this section, we consider separation results between applicative theories.
Following the proof of \cite[Lemma $6$]{strahm:03}, one can show that, together with the theory {\sf B}, $(\SSIl)$ entails $(\SSI)$ and $(\SSIl)^\vee$ entails $(\SSI)^\vee$ (all the auxiliary functions used by Strahm are in $\bff{FP}$ and thus provably total in {\sf CHT} and {\sf PST}). 
Since $\bff{TC}^0\subseteq\bff{CH}$ and $\bff{NC}^1\subseteq\bff{PSPACE}$, we thus obtain the following two proper inclusions between applicative theories:
\begin{enumerate}
    \item ${\sf LCHT}\subsetneq {\sf CHT}$
    \item ${\sf ALT}\subsetneq {\sf PST}$.
\end{enumerate}
The separation results between $\bff{TC}^0$ and $\bff{NC}^1$, between $\bff{NC}^1$ and $\bff{CH}$, and between $\bff{CH}$ and $\bff{PSPACE}$ are open problems.
 
Here, we show that oracle constructions allow one to separate applicative theories, even if they are not relativized.
More precisely, we show that {\sf ALT} cannot prove the induction axiom $(\SSIl)$.
The proof relies on a series of remarks, following a similar result in \cite{ertel:26}.
Define $\mathtt{sum} = \mathtt{rec}\,r$, where
$r = \lambda fgyx.\,\ms{c_\epsilon}y(g\eps x)(\ms{c_W}(gyx)(fg(\ms{p_\ell}y)x)
(\ms{s_\ell}(fg(\ms{p_\ell}y)x))(\ms{s_\ell}(fg(\ms{p_\ell}y)x)))$.
For $x,y\in\ms{W}$, and $g:\ms{W}^2\to\{\eps,0\}$, $\mathtt{sum} \, gyx$ returns the number of words $\eps\leq_\ell z\leq_\ell y$ for which $gzx$ ends in $0$.
Define the formula ${\sf Fun}(\mathtt{sum})$:
    $(\forall g)\big((g:\ms{W}^2\to\ms{W})\,\to\,(\mathtt{sum}\,g:\ms{W}^2\to\ms{W})\big)$.
Clearly, ${\sf Fun}(\mathtt{sum})$ ``reasons'' about a \textit{functional}: it states that, whenever $g$ represents a total function, so does $\mathtt{sum}\,g$.
A series of remarks follows:

\begin{proposition}\label{prop:CHT_functional}
{\sf CHT} can prove ${\sf Fun}(\mathtt{sum})$.
\end{proposition}
\begin{proof}
We argue in {\sf CHT}.
Assume $g:\ms{W}^2\to\ms{W}$. 
We use induction on the formula $A(y,x,z):= \mathtt{sum}\,gyx=z$.
Fix the parameter $x\in\ms{W}$.
For $y=\epsilon$, $\mathtt{sum}\,g\epsilon x = g\eps x$, so $\exists z\in\ms{W}.A(\epsilon,x,z)$ holds.
Now, let $y\in\ms{W}$ and $z$ be arbitrary and assume that $A(y,x,z)$ holds, i.e., $\mathtt{sum} gyx=z$. 
Since $\mathtt{sum}\,g(\ms{s_\ell}y)x$ is now $z$ or $\ms{s_\ell}z$, the conclusion of the induction step follows.
By $(\SSIl)$ it follows that $\forall y,x\in\ms{W}.\,\exists z\in\ms{W}.\,A(y,x,z)$, and from this it follows that $\mathtt{sum}\, g\,:\,\ms{W}^2\to\ms{W}$.
\end{proof}

\noindent
The interesting aspect of this proof is that it shows that the axiom scheme $(\SSIl)$ can act as a black box, ``counting'' independently of the structure of the function $g$.

\begin{theorem}[\cite{baker:75}]\label{thm:oracle_separation_baker}
There is a recursive oracle $A$ such that $\bff{P}^A\subsetneq\bff{NP}^A$.
\end{theorem}
\noindent
Since $\bff{FNC}^1\subseteq\bff{FP}$ and $\bff{NP}\subseteq\bff{FCH}$, this implies, in particular, that the same oracle separates $\bff{FNC}^1$ from $\bff{FCH}$.
Let $A$ be the said oracle and denote by ${\sf ALT}+A$ the applicative theory ${\sf ALT}$ extended with the axiom $\ms{W}(x)\to \ms{W}(t_A(x))$, where $t_A(x)$ is a given $\cal{L}$-term representing the characteristic function of $A$ (Def. \ref{def:provably_total_function}).\footnote{We relativize $\bff{FP}$ instead of $\bff{FNC}^1$, as $\bff{FP}$ has a much more robust relativized model.}

\begin{proposition}\label{prop:oracle_in_theory}
$\comp{{\sf ALT}+A}\subseteq \bff{FP}^A$.
\end{proposition}
\begin{proof}
Each axiom and rule of {\sf ALT} can be realized by an $\bff{FNC}^1$ function and thus, in particular, by an $\bff{FP}$ function.
The extra axiom is $\ms{W}(x)\To\ms{W}(1_A(x))$, which is realized by the $\bff{FP}^A$ function $F(\rho) = \pair{0,1_A(\rho)}$.
Now, the statement follows by noticing that the realizability theorems \textit{relativize}, as they consist of simple simulations of the realizing functions given by induction (see App.~\ref{app:oracle_in_theory} for an example). 
\end{proof}

\noindent
We finally conclude by proving the following theorem.

\begin{theorem}\label{thm:separation}
${\sf ALT} \neq {\sf CHT}$. In particular, {\sf ALT} cannot prove $(\SSIl)$.
\end{theorem}
\begin{proof}
We show that {\sf ALT} cannot prove $(\SSIl)$, by showing that it cannot prove ${\sf Fun}(\mathtt{sum})$.
Recall that the language witnessing the separation between $\bff{P}^A$ and $\bff{NP}^A$ is $L(A) = \{x\,|\,\exists z\in\{0,1\}^{|x|}\cap A\}$ \cite[Sec. $3$]{baker:75}.
Suppose that ${\sf ALT}\vdash{\sf Fun}(\mathtt{sum})$.
Define a $\cal{L}$-term $g$ representing the recursive function $G(y,x) = 1_A(y)\wedge |x|=|y|$ (outputting $\eps$ and $0$, instead of $0$ and $1$).
From the axiom $\ms{W}(x)\to\ms{W}(t_Ax)$, ${\sf ALT}+A$ can prove the totality of $g$.
Thus, from $g:\ms{W}^2\to\ms{W}$ and ${\sf Fun}(\mathtt{sum})$, ${\sf ALT}+A$ proves $\mathtt{sum}\, g:\ms{W}^2\to\ms{W}$ and thus the totality of $f = \lambda x.\mathtt{sum} g(\ms{1}\ms{\times}x)x$.
It follows that ${\sf ALT}+A$ proves the totality of the characteristic function of $L(A)$, so $1_{L_A}\in\comp{ALT+A}\subseteq\bff{FP}^A$.
This contradicts Th.~\ref{thm:oracle_separation_baker}.
\end{proof}

\noindent
Unlike the separations mentioned at the start of this section, it is not obvious whether the above result establishes a strict containment, or an incomparability of theories. That is, can ${\sf CHT}$ prove $(\SSI)^\vee$? If so, we would have an interesting tradeoff between logical complexity of the induction formulas and the length of the induction process.
\section{Conclusion}\label{sec:conclusion}
We introduced a family of axiom schemes, \emph{naturally} tailored for counting, which yields, under small variations, proof-theoretic characterizations of $\mathbf{FTC}^0, \mathbf{FNC}^1, \mathbf{FCH}$, and $\mathbf{FPSPACE}$.
Notably, for $\mathbf{FTC}^0$ and $\mathbf{FCH}$, no corresponding applicative theories were defined before.
The simplicity of these schemes is a promising starting point to investigate other classes involving different forms of counting, such as counting modulo $p$ (which characterizes $\bff{FAC}^0$ with modulo $p$ gates, see~\cite{clotetakeuti,ICALP}). These classes have turned out to be especially relevant to circuit complexity, being some of the few for which separation results are known~\cite{Furst,ajtai,smolensky:87}.

Moreover, we simplify the known characterization of $\mathbf{FAC}^0$~\cite{eberhard:15}, by requiring the variable $z$ in $(\BSI)$ to range over the entire universe, instead of introducing an extra predicate $\mathsf{V}$. Intuitively, our theory must prove the step formula without ``knowing the value'' of $z$.
By generalizing a remark from~\cite{eberhard:15}, we showed that disjunctions in lexicographic induction schemes allow one to represent a form of constant-bounded recursion, thus capturing $\bff{FPSPACE}$ and solving an open question posed in~\cite[Q.~21]{kahle:11}: using the same idea, allowing disjunctions in Kahle and Oitavem's monotone induction 
yields a characterization of $\mathbf{FPSPACE}$. 
This observation suggests a natural candidate for an applicative theory of $\mathbf{ALINTIME}$, using the recursion-theoretic characterization by Clote~\cite{clote:97} (and~\cite[Th.~3.59]{clote:99}).%

Finally, we establish some of the first separation results in the applicative setting.
Intuitively, applicative theories should be easier to separate than their complexity classes, as they can express much more complex formulas than simply totality statements.
We illustrate this for {\sf ALT} and {\sf CHT}. The given oracle-based separation technique readily adapts to theories with known relativized class separations (e.g.,~any of those in~\cite{strahm:03,kahle:11,eberhard:15}).
For classes for which an oracle separation is not known, such as $\mathbf{FCH}$ and $\mathbf{FPSPACE}$, a possible solution might be to study them on even more complex formulas than functional totality statements.
In general, we are interested in how relationships between complexity classes map onto their respective applicative theories, which are able to represent higher-order versions of them \cite{strahm:04}.
For example, a natural question arising in this setting is whether the implication $\mathbf{FTC}^0 = \mathbf{FNC}^1 \to \mathbf{FCH} = \mathbf{FPSPACE}$ holds on fragments of their corresponding theories. That is, whether 
${\sf LCHT} = {\sf ALT} \to {\sf CHT} = {\sf PST}$ holds for given classes of formulas.
\newpage

\section*{Declaration on Generative AI}
The authors used Mistral's Vibe to check grammar and proof details. The authors take full responsibility for the publication's content.

\bibliography{bib}

@article{ishihara:99,
  title={Function algebraic characterizations of the polytime functions},
  author={Ishihara, Hajime},
  journal={computational complexity},
  volume={8},
  number={4},
  pages={346--356},
  year={1999},
  publisher={Springer}
}

@article{barrington:90,
  title={On uniformity within NC1},
  author={Barrington, David A Mix and Immerman, Neil and Straubing, Howard},
  journal={Journal of Computer and System Sciences},
  volume={41},
  number={3},
  pages={274--306},
  year={1990},
  publisher={Elsevier}
}

@article{ruzzo:81,
  title={On uniform circuit complexity},
  author={Ruzzo, Walter L},
  journal={Journal of Computer and System Sciences},
  volume={22},
  number={3},
  pages={365--383},
  year={1981},
  publisher={Elsevier}
}

@inproceedings{stearns:65,
  title={Hierarchies of memory limited computations},
  author={Stearns, Richard Edwin and Hartmanis, Juris and Lewis, Philip M},
  booktitle={6th annual symposium on switching circuit theory and logical design (SWCT 1965)},
  pages={179--190},
  year={1965},
  organization={IEEE}
}

@article{borodin:77,
  title={On relating time and space to size and depth},
  author={Borodin, Allan},
  journal={SIAM journal on computing},
  volume={6},
  number={4},
  pages={733--744},
  year={1977},
  publisher={SIAM}
}

@article{caussinus:98,
  title={NondeterministicNC1Computation},
  author={Caussinus, Herv{\'e} and McKenzie, Pierre and Th{\'e}rien, Denis and Vollmer, Heribert},
  journal={Journal of Computer and System Sciences},
  volume={57},
  number={2},
  pages={200--212},
  year={1998},
  publisher={Elsevier}
}

@article{allender:99,
  title={The permanent requires large uniform threshold circuits},
  author={Allender, Eric},
  journal={Chicago Journal of Theoretical Computer Science},
  pages={1--19},
  year={1999},
  publisher={Mit Press}
}

@article{ertel:26,
  author = {Marcel Ertel and Eduardo Skapinakis},
  title = {Non-deterministic operations in applicative theories},
  journal = {Manuscript},
  year = {2026}
}

@book{balcazar:88,
  author = {Balc{\'a}zar, Jos{\'e} L and D{\'\i}az, Josep and Gabarr{\'o}, Joaquim},
  title = {Structural complexity I},
  publisher = {Springer Science \& Business Media},
  year = {1988}
}

@article{baker:75,
  author = {Baker, Theodore and Gill, John and Solovay, Robert},
  title = {Relativizations of the {P}=?{NP} question},
  journal = {SIAM Journal on computing},
  volume = {4},
  number = {4},
  pages = {431--442},
  year = {1975}
}

@article{aehlig:16,
  title={Relativizing small complexity classes and their theories},
  author={Aehlig, Klaus and Cook, Stephen and Nguyen, Phuong},
  journal={Computational complexity},
  volume={25},
  number={1},
  pages={177--215},
  year={2016},
  publisher={Springer}
}

@article{bellantoni:95,
  author = {Bellantoni, Stephen},
  title = {Predicative recursion and the polytime hierarchy},
  journal = {Feasible Mathematics II},
  pages = {15--29},
  year = {1995}
}

@book{beeson:85,
  title={Foundations of constructive mathematics: Metamathematical studies},
  author={Beeson, Michael J},
  year={1985},
  publisher={Springer Science \& Business Media}
}

@article{jager:95,
  title={Totality in applicative theories},
  author={J{\"a}ger, Gerhard and Strahm, Thomas},
  journal={Annals of Pure and Applied Logic},
  volume={74},
  number={2},
  pages={105--120},
  year={1995},
  publisher={Elsevier}
}

@incollection{feferman:79,
  title = {Constructive Theories of Functions and Classes},
  editor = {Maurice Boffa and Dirkvan Dalen and Kenneth Mcaloon},
    series = {Studies in Logic and the Foundations of Mathematics},
  publisher = {Elsevier},
  volume = {97},
  pages = {159-224},
  year = {1979},
  booktitle = {Logic Colloquium '78},
  author = {Solomon Feferman}
}

@article{toran:91,
  author = {Torán, Jacobo},
  year = {1991},
  month = {07},
  pages = {753-774},
  title = {Complexity Classes Defined by Counting Quantifiers},
  volume = {38},
  journal = {Journal of the ACM}
}

@article{clote:97,
  title={Nondeterministic stack register machines},
  author={Clote, Peter},
  journal={Theoretical computer science},
  volume={178},
  number={1-2},
  pages={37--76},
  year={1997},
  publisher={Elsevier}
}

@incollection{clote:93,
    author = {Clote, Peter},
    editor = {Clote, Peter and ČEk, Jan KrajÍ},
    title = {On polynomial size Frege proofs of certain combinatorial principles},
    booktitle = {Arithmetic, proof theory, and computational complexity},
    publisher = {Oxford University Press},
    year = {1993},
    pages = {162 -- 184}
}

@book{buss:85,
  author = {Buss, Samuel R},
  title = {Bounded arithmetic},
  publisher = {Princeton University},
  year = {1985}
}

@phdthesis{calamai:08,
  author = {Calamai, Giacomo},
  title = {Proof-theoretic contributions to computational complexity},
  school = {University of Siena},
  year = {2008}
}

@article{cantini:02,
  author = {Cantini, Andrea},
  title = {Polytime, combinatory logic and positive safe induction},
  journal = {Archive for Mathematical Logic},
  volume = {41},
  number = {2},
  pages = {169--189},
  year = {2002}
}

@article{cantini:05,
  author = {Cantini, Andrea},
  title = {Remarks on applicative theories},
  journal = {Annals of Pure and Applied Logic},
  volume = {136},
  number = {1-2},
  pages = {91--115},
  year = {2005}
}

@article{clote:99,
  author = {Clote, Peter},
  title = {Computation models and function algebras},
  journal = {Studies in Logic and the Foundations of Mathematics},
  volume = {140},
  pages = {589--681},
  year = {1999}
}

@inproceedings{cobham:64,
  author = {Cobham, Alan},
  title = {The intrinsic computational difficulty of functions},
  booktitle = {International Congress for Logic, Methodology, and the Philosophy of Science},
  pages = {24--40},
  year = {1964}
}

@article{eberhard:15,
  author = {Eberhard, Sebastian},
  title = {Applicative theories for logarithmic complexity classes},
  journal = {Theoretical Computer Science},
  volume = {585},
  pages = {115--135},
  year = {2015}
}

@article{feferman:75,
  author = {Feferman, Solomon},
  title = {A language and axioms for explicit mathematics},
  journal = {Algebra and Logic, Lecture Notes in Mathematics},
  volume = {450},
  pages = {87--139},
  year = {1975}
}

@article{gentzen:35,
  author = {Gentzen, Gerhard},
  title = {Untersuchungen {\"u}ber das logische schlie{\ss}en. {I}},
  journal = {Mathematische Zeitschrift},
  volume = {39},
  pages = {176--210},
  year = {1935}
}

@inproceedings{kahle:11,
  title={An applicative theory for {FPH}},
  author={Kahle, Reinhard and Oitavem, Isabel},
  booktitle={S. van Bakel, S. Berardi, U. Berger (Eds.), Classical Logic and Computation 2010 in: Electronic Proceedings of Theoretic Computer Science, vol. 47},
  year={2011},
  pages = {44--56}
}

@article{buss:92,
  title={The Graph of Multiplication is Equivalent to Counting},
  author={Samuel R. Buss},
  journal={Inf. Process. Lett.},
  year={1992},
  volume={41},
  pages={199-201}
}

@article{strahm:04,
author = {Strahm, Thomas},
year = {2004},
month = {12},
pages = {159-176},
title = {A proof-theoretic characterization of the basic feasible functional},
volume = {329},
journal = {Theoretical Computer Science}
}

@article{allender:90,
  title = {Counting hierarchies: polynomial time and constant},
  journal = {Bull. EATCS},
  volume = {40},
  pages = {182-194},
  year = {1990},
 author = {Allender, Eric and Wagner, Klaus W.}
}

@article{valiant:79,
  title={The complexity of computing the permanent},
  author={Valiant, Leslie G},
  journal={Theoretical computer science},
  volume={8},
  number={2},
  pages={189--201},
  year={1979},
  publisher={Elsevier}
}

@article{wagner:86,
  title={The complexity of combinatorial problems with succinct input representation},
  author={Wagner, Klaus W},
  journal={Acta informatica},
  volume={23},
  number={3},
  pages={325--356},
  year={1986},
  publisher={Springer}
}

@article{vollmer:96,
  title={Recursion theoretic characterizations of complexity classes of counting functions},
  author={Vollmer, Heribert and Wagner, Klaus W},
  journal={Theoretical Computer Science},
  volume={163},
  number={1-2},
  pages={245--258},
  year={1996},
  publisher={Elsevier}
}

@article{chandra:84,
    author = {Chandra AK, Stockmeyer L, Vishkin U},
    title = {Constant depth reducibility},
    journal = {SIAM Journal on Computing},
    year = {1984},
    volume = {13},
    number = {2},
    pages = {423--439}
}

@article{lago:22,
  author = {Dal Lago, Ugo and Kahle, Reinhard and Oitavem, Isabel},
  title = {Implicit recursion-theoretic characterizations of counting classes},
  journal = {Archive for Mathematical Logic},
  volume = {61},
  number = {7},
  pages = {1129--1144},
  year = {2022}
}

@inproceedings{leivant:94,
  author = {Leivant, Daniel and Marion Jean-Yves},
  title = {Ramified recurrence and computational complexity II: substitution and poly-space},
  booktitle = {International Workshop on Computer Science Logic},
  publisher = {Springer Berlin Heidelberg},
  year = {1994}
}

@article{oitavem:22,
  author = {Oitavem, Isabel},
  title = {The polynomial hierarchy of functions and its levels},
  journal = {Theoretical Computer Science},
  volume = {900},
  pages = {25--34},
  year = {2022}
}

@phdthesis{strahm:96,
  author = {Strahm, Thomas},
  title = {On the proof theory of applicative theories},
  school = {Universität Bern},
  year = {1996}
}

@article{strahm:97,
  author = {Strahm, Thomas},
  title = {Polynomial time operations in explicit mathematics},
  journal = {The Journal of Symbolic Logic},
  volume = {62},
  number = {2},
  pages = {575--594},
  year = {1997}
}

@article{strahm:03,
  author = {Strahm, Thomas},
  title = {Theories with self-application and computational complexity},
  journal = {Information and Computation},
  volume = {185},
  number = {2},
  pages = {263--297},
  year = {2003}
}

@inproceedings{strahm:10,
  author = {Strahm, Thomas},
  title = {Weak theories of operations and types},
  booktitle = {R. Schindler (Ed.), Ways of Proof Theory},
  pages = {441--468},
  year = {2010}
}

@book{troelstra:00,
  author = {Troelstra, Anne S. and Helmut Schwichtenberg},
  title = {Basic proof theory},
  publisher = {Cambridge University Press},
  year = {2000}
}

@article{parikh, title={Existence and feasibility in arithmetic}, volume={36}, number={3}, journal={Journal of Symbolic Logic}, author={Parikh, Rohit}, year={1971}, pages={494–508}}

@book{vollmer, author={Vollmer, Heribert}, title={Introduction to Circuit Complexity: A Uniform Approach}, publisher={Springer}, year={1999}}

@InProceedings{clotetakeuti,
author={Clote, Peter
and Takeuti, Gaisi},
editor={Clote, Peter
and Remmel, Jeffrey B.},
title={First Order Bounded Arithmetic and Small Boolean Circuit Complexity Classes},
booktitle={Feasible Mathematics II},
year={1995},
publisher={Birkh{\"a}user Boston},
address={Boston, MA},
pages={154--218}}

@article{Furst,
  title={Parity, circuits, and the polynomial-time hierarchy},
  author={Merrick L. Furst and James B. Saxe and Michael Sipser},
  journal={Mathematical systems theory},
  year={1981},
  volume={17},
  pages={13-27}
}

@inproceedings{ICALP,
  author       = {Melissa Antonelli and
                  Arnaud Durand and
                  Rui Li},
  editor       = {Sayan Bhattacharya and
                  Danupon Nanongkai and
                  Michael Benedikt and
                  Gabriele Puppis},
  title        = {Recursion and Proof Theoretical Characterizations of Small Circuit
                  Classes with Modulo Counting via Discrete Differential Equations},
  booktitle    = {53rd International Colloquium on Automata, Languages, and Programming,
                  {ICALP} 2026, Royal Holloway, University of London, Egham, United
                  Kingdom, July 7-10, 2026},
  series       = {LIPIcs},
  volume       = {374},
  pages        = {162:1--162:19},
  publisher    = {Schloss Dagstuhl - Leibniz-Zentrum f{\"{u}}r Informatik},
  year         = {2026}
}

@book{Immerman99,
  author       = {Immerman, Neil},
  title        = {Descriptive Complexity},
  publisher    = {Springer},
  year         = {1999}
}

@article{BarringtonImmermanStraubing,
    author = {Barrington, David A.M. and Immerman, Neil and Straubing, Howard},
    title = {On uniformity within \text{NC}$^1$},    
    journal = {J. of Comput. and Syst. Sc.},
    volume = {41},
    pages = {274--306},
    year = {1990}
}

@article{GurevichLewis,
    author = {Gurevich, Yuri and Lewis, H.},
    title = {A logic for constant-depth circuit},
    journal = {Inf. Control},
    volume = {61},
    year = {1984},
    pages = {65--74}
}

@inproceedings{Lindell,
    author = {Lindell, Steven},
    title = {A purely logical characterization of circuit uniformity},
    booktitle = {7th Structure in Complexity Theory Conf.},
    year = {1992},
pages = {185--192}
}

@article{ComptonLaflamme,
  author       = {Compton, Kevin J. and Laflamme, Compton},
  title        = {An Algebra and a Logic for {NC\({^1}\)}},
  journal      = {Inf. Comput.},
  volume       = {87},
  number       = {1/2},
  pages        = {240--262},
  year         = {1990}
}

@article{Allen,
  author = {Allen, Bill},
  year = {1991},
  title = {Arithmetizing Uniform {NC}},
  journal = {Annals of Pure and Applied Logic},
  volume = {53},
  pages = {1-50}
}

@article{CookMorioka,
  author = {Cook, Stephen and Morioka,  Tsuyoshi},
  year = {2005},
  title = {Quantified propositional calculus and a second-order theory for \text{NC}$^1$},
  journal = { Archive for Mathematical Logic},
  volume = {44},
  pages = {711-749}
}

@phdthesis{Bennett,
    author = {Bennett, James H.},
    title = {On Spectra},
    school = {Princeton University},
    year = {1962}
}

@article{BellantoniCook,
    author = {Bellantoni, Stephen and Cook, Stephen},
    title = {A New Recursion-Theoretic Characterization of Poly-Time Functions},
    journal = {Computational Complexity},
    volume = {2},
    year = {1992},
    pages = {97--110}
}

@article{CookNguyen,
	author = {Cook, Stephen and Nguyen, Phuong},
	journal = {{Log. Meth. Comput. Sci.}},
	number = {1--40},
	title = {Theories for {TC}$^0$ and other small circuit classes},
	volume = {2},
	year = {2006}}

@article{Arai,
    author = {Arai, Toshiyasu},  
    title = {A bounded arithmetic {AID} for \text{F}rege systems},
    journal = {Annals of Pure amd Applied Logic},
    volume = {103},
    pages = {155-199},
    year = {2000}
}

@inproceedings{smolensky:87,
  title={Algebraic methods in the theory of lower bounds for Boolean circuit complexity},
  author={Smolensky, Roman},
  booktitle={Proceedings of the nineteenth annual ACM symposium on Theory of computing},
  pages={77--82},
  year={1987}
}

@article{ajtai,
  title={Sigma$^1_1$-Formulae on finite structures},
  author={Mikl{\'o}s Ajtai},
  journal={Annals of Pure and Applied Logic},
  year={1983},
  volume={24},
  pages={1-48},
}

@inproceedings{ADK24a,
  author       = {Melissa Antonelli and
                  Arnaud Durand and
                  Juha Kontinen},
  editor       = {Rastislav Kr{\'{a}}lovic and
                  Anton{\'{\i}}n Kucera},
  title        = {A New Characterization of {FAC}$^0$ via Discrete Ordinary
                  Differential Equations},
  booktitle    = {49th International Symposium on Mathematical Foundations of Computer
                  Science, {MFCS} 2024, Bratislava, Slovakia, August 26-30, 2024},
  series       = {LIPIcs},
  volume       = {306},
  pages        = {10:1--10:18},
  publisher    = {Schloss Dagstuhl - Leibniz-Zentrum f{\"{u}}r Informatik},
  year         = {2024}
}

@techreport{Clote88,
    author = {Clote, Peter},
    title = {A Sequential Characterization of the Parallel Complexity Class {NC}},
    institution = {Boston College},
    year = {1988}
}

@incollection{Clote1990, 
address={Boston, MA}, 
series={Progress in Computer Science and Applied Logic}, 
title={Sequential, machine-independent characterizations of the parallel complexity classes AlogTIME, AC$^k$, NC$^k$ and NC}, 
 booktitle={Feasible Mathematics: A Mathematical Sciences Institute Workshop, Ithaca, New York, June 1989}, 
publisher={Birkhäuser}, 
author={Clote, Peter}, 
editor={Buss, Samuel R. and Scott, Philip J.}, 
year={1990}, 
pages={49-69} 
}

@article{oitavem:11,
  author       = {Isabel Oitavem},
  title        = {A recursion-theoretic approach to {NP}},
  journal      = {Annals of Pure and Applied Logic},
  volume       = {162},
  number       = {8},
  pages        = {661--666},
  year         = {2011}
}

@inproceedings{dal:21,
  author       = {Ugo Dal Lago and
                  Reinhard Kahle and
                  Isabel Oitavem},
  editor       = {Filippo Bonchi and
                  Simon J. Puglisi},
  title        = {A Recursion-Theoretic Characterization of the Probabilistic Class
                  {PP}},
  booktitle    = {46th International Symposium on Mathematical Foundations of Computer
                  Science, {MFCS} 2021, Tallinn, Estonia, August 23-27, 2021},
  series       = {LIPIcs},
  volume       = {202},
  pages        = {35:1--35:12},
  publisher    = {Schloss Dagstuhl - Leibniz-Zentrum f{\"{u}}r Informatik},
  year         = {2021}
}

@article{dal:22,
  author       = {Ugo Dal Lago and
                  Reinhard Kahle and
                  Isabel Oitavem},
  title        = {Implicit recursion-theoretic characterizations of counting classes},
  journal      = {Arch. Math. Log.},
  volume       = {61},
  number       = {7-8},
  pages        = {1129--1144},
  year         = {2022}
}

@article{Gill77,
	author = {Gill, J.T.},
	journal = {SIAM Journal on Computing},
	number = {4},
	pages = {675--695},
	title = {{Computational Complexity of Probabilistic Turing Machines}},
	volume = {6},
	year = {1977}}

@inproceedings{Toda89,
	author = {Toda, Seinosuke},
	booktitle = {30th Annual Symposium on Foundations of Computer Science},
	doi = {10.1109/SFCS.1989.63527},
	pages = {514--519},
	title = {{On the Computational Power of PP and $\sharp$P}},
	year = {1989}}

@article{Toda91,
	author = {Toda, Seinosuke},
	doi = {https://doi.org/10.1137/0220053},
	journal = {SIAM Journal on Computing},
	number = {5},
	pages = {865--877},
	title = {{PP is as Hard as the Polynomial-Time Hierarchy}},
	volume = {20},
	year = {1991}}

@inproceedings{Toran88,
	author = {Tor\'an, Jacobo},
	booktitle = {Proceedings. Structure in Complexity Theory Third Annual Conference},
	pages = {213--223},
	title = {{An Oracle Characterization of the Counting Hierarchy}},
	year = {1988}}

@phdthesis{Simon75,
    author = {Janos Simon},
    title = {On Some Centrall Problems in Computational Complexity},
    school = {Cornell University},
    year = {1975}
}

@article{Santos69,
 author = {Eugene S. Santos},
 journal = {Proceedings of the American Mathematical Society},
 number = {3},
 pages = {704--710},
 publisher = {American Mathematical Society},
 title = {Probabilistic {T}uring Machines and Computability},
 urldate = {2025-12-27},
 volume = {22},
 year = {1969}
}

\appendix

\section{Preliminaries}\label{app:preliminaries}

\subsection{On counting classes and their characterizations}\label{sec:PHCH}

Boolean circuits are vertex-labelled directed acyclic graphs whose nodes are 
either input or output nodes or are labelled with a Boolean function  from the  set $\{\wedge, \vee, \neg\}$.
A Boolean circuit with majority gates allows in addition gates labelled by the function \textsc{Maj}, that  outputs 1 when the majority of its inputs are 1's.
A family of circuits $(C_n)$ is $\mathbf{DLOGTIME}$-uniform if there is a Turing machine (with a random access tape) that decides in deterministic logarithmic time the corresponding direct connection language of the circuit.
%
When dealing with circuits, the resources of interest are \emph{size}, i.e.~the number of their gates, and \emph{depth}, i.e.~the length of the longest path from the input to the output, leading to the definition of the classes below.

\begin{definition}
\label{def:ACandTC}
For $i\in \Nat$, the class $\mathbf{AC}^i$ (resp., $\mathbf{NC}^i$) is the class of languages recognized by a $\mathbf{DLOGTIME}$-uniform family of unbounded (resp. bounded) Boolean circuits of polynomial size and depth $O((\log n)^i)$.
If  circuits also include \textsc{Maj} gates, the corresponding class is called $\mathbf{TC}^i$, while if \textsc{Mod $p$} gates (for $p\in \mathbb{N}$), i.e.~gates returning the sum of the inputs modulo $p$, are added, we obtain the class $\mathbf{ACC[p]}$
We denote by $\mathbf{FAC}^i$ and $\mathbf{FTC}^i$ the corresponding function classes.
\end{definition}
\noindent
Notably, some of the very few separation results in circuit complexity are precisely established considering corresponding counting classes.
That Parity is not doable by standard Boolean circuits of polynomial size and constant depth~\cite{Furst,ajtai} implies the separation between $\mathbf{FAC}^0$ and $\mathbf{FACC[2]}$. For any level above $\mathbf{FNC}^1$, it is not known whether inclusions for levels in the circuit hierarchies are strict:
$$
\mathbf{FAC}^0 \subsetneq \mathbf{FACC[2]} \subsetneq \mathbf{FTC}^0 \subseteq \mathbf{FNC}^1 \subseteq \mathbf{FAC}^1 \subseteq \mathbf{FNC}^2 \dots
$$

Moving beyond, we will follow the convention below to define oracle access.

\begin{notation}
We use $M^O$ to denote a Turing machine (TM) accessing an oracle set $O\subseteq \Nat$
and $\#\acc_{M,x}$ 
for the number of accepting computation paths of $M$ on input $x$.
\end{notation}

The polynomial-time hierarchy ($\PH$) is the complexity analogue of the arithmetical hierarchy. 
It generalizes the well-known classes $\NP$ and $\coNP$ by successively relativizing non-deterministic TMs.
As a class of predicates (characteristic functions), the counting hierarchy ($\cc{CH}$) has historically been defined both in terms of counting operator alternation~\cite{wagner:86} and via oracles~\cite{Toran88,toran:91}, considering building blocks defined in terms of either probabilistic TMs~\cite{Santos69,Gill77} or counting machines~\cite{Simon75,valiant:79}.

\begin{definition}
A function is in $\sP$ iff it can be computed by  a poly-time counting machine, i.e.~by an NTM returning the number of its accepting computation paths written in binary.
\end{definition}
Just as $\prob{SAT}$ is the canonical complete problem for $\cc{NP}$, the canonical complete function for $\sP$ is the function computing the number of satisfying assignments of a Boolean formula~\cite{valiant:79}.

\begin{definition}\label{def:complexity_classes}
    The polynomial and counting hierarchies are defined respectively as $\cc{PH} := \bigcup_{n\geq0}\Sigma_n^p = \bigcup_{n\geq0}\Pi_n^p$ and $\cc{FCH} := \bigcup_{n\geq0}\cc{FCH}_n$, where, for $n\ge 1$:
    \begin{align*}
    \Sigma^p_0 &:= \Pi^p_0 := \cc{P} \quad \quad \quad \Sigma^p_{n+1} := \NP^{\Sigma^p_n} \quad \quad \quad \Pi^p_{n+1} := \coNP^{\Sigma^p_n} \\
    \cc{FCH}_0 &:= \bff{FP} \quad \quad \quad \quad \; \;
    \cc{FCH}_{n} := \sP^{\cc{FCH}_{n-1}}.
    \end{align*}
\end{definition}

\noindent
These classes are related by a fundamental result due to Toda~\cite{Toda89,Toda91}:
$\cc{PH}\subseteq\cc{P}^{\sP}$.

\subsection{On recursion-theoretic characterizations}\label{sec:algebras}
A prominent approach to the study of computability and complexity involves function algebras, i.e.,~sets of functions closed under specific operations:
$[\mathcal{B}; \mathcal{O}_1, \dots, \mathcal{O}_k]$.
In computability theory, the principal operations are composition and primitive recursion; since the 1960s, restrictions on the latter have been used to characterize complexity classes in a machine-independent way~\cite{cobham:64}.

One of the foundational results in this area is Cobham's characterization of $\FP$ by an algebra whose defining scheme is Bounded Recursion on Notation ($\mathsf{BRN}$), due to Bennett~\cite{Bennett}:
a function $F$ is defined by $\mathsf{BRN}$ from $G$, $H_0$, $H_1$ and $K$ if $F(yi, x)=H_i(y, x,F(y,x))$, for $i\in\{0,1\}$, and for any $x$ and $y$, $F(y,x) \leq K(y,x)$.
To avoid the \textit{explicit} bound on the function size, alternative machine-independent approaches to $\mathbf{FP}$ emerged, based on input sorting~\cite{BellantoniCook}, data tiering~\cite{leivant:94}, and restrictions on the allowed step functions~\cite{ishihara:99}.
These methodologies led to a host of characterizations of complexity classes based on different recursion-theoretic disciplines (see \cite{clote:99} for an overview).

In 1988/90, Clote introduced the algebra $\mathcal{A}_0$ to capture functions in the log-time hierarchy (equivalent to $\AC^0$) using so-called \emph{concatenation recursion on notation} ($\mathsf{CRN}$)~\cite{Clote88,Clote1990}.
In particular, it relies on the scheme of concatenation recursion on notation
($\mathsf{CRN}$).\footnote{Clote's original definitions are formulated over $\mathbb{N}$
(see also~\cite{clote:99}); here and below we transpose them to $\bb{W}$ (Notation~\ref{not:notation}).}
A function $F$ is defined by \emph{concatenation recursion on notation} from $G$ and $H_i$,
with $i\in\{0,1\}$, if for all $x,y$: 
    \begin{align*}
        F(\eps,x) = G(x) \ \mbox{ and }  \quad  
        F(yi, x) &= S_{H_i(y,x)}(F(y,x)) 
    \end{align*}
Clote's function algebra 
is defined as the class:
$$
\mathcal{A}_0 = [0, \pi^p_i, S_0, S_1, \ell, \text{BIT}, \#; \circ, \mathsf{CRN}],
$$
where BIT$(y,x)$ returns the $y^{th}$ bit of $x$, and $y\# x= 2^{\ell(y)\times \ell(x)}$ is the smash function.
The larger classes $\bff{FAC}^i$ and $\bff{FNC}$ are captured adding \emph{weak bounded recursion on notation}~\cite{Clote1990}, while
$\bff{FNC}^1$ has been characterized both in terms of the \emph{$k$-bounded recursion on notation} ($k$-$\mathsf{BRN}$) scheme~\cite{clote:93} and relying on the basic function $tree$~\cite{Clote1990}, where, for any $k\ge 4$, a function $F$ is said to be defined by $k$-$\mathsf{BRN}$ from $G$ and $H$ if $F(\eps,x)=G(x)$ and $F(yi,x)=H_i(y,x,F(y,x))$, for $F$ taking only values in $\{0,\dots, k\}$.
In 1990, an alternative algebra (and logic), based on a scheme called \emph{upward tree recursion}, was introduced to characterize $\mathbf{FNC}^1$~\cite{ComptonLaflamme}.
In 1991, a new recursion-theoretic characterization (and arithmetic \emph{\`a la Buss}) for $\mathbf{FNC}$ was presented~\cite{Allen}.
In 1992 and 1995, other small circuit classes, including $\mathbf{FTC}^0$, were considered (and re-considered) by Clote and Takeuti, who also introduced bounded theories for them~\cite{clotetakeuti}.
%
%
In~\cite{ADK24a}, the first ODE-based characterizations for $\mathbf{FAC}^0$ and $\mathbf{FTC}^0$ were introduced. 
Related approaches to capture small circuit classes have also been provided in the framework of model- and proof-theory, both in the first and second order~\cite{Immerman99,BarringtonImmermanStraubing,GurevichLewis,Lindell,ComptonLaflamme,Allen,clotetakeuti,CookNguyen,Arai,CookMorioka}.

Regarding counting classes, characterizations in the style of Bellantoni and Cook have been developed for $\PH$~\cite{bellantoni:95,oitavem:11,oitavem:22}, $\PP$~\cite{dal:21}, as well as for $\sP$ and $\FCH$~\cite{dal:22}.
Furthermore, counting classes have been characterized in~\cite{vollmer:96} by the schemes $\WProd$ and $\Sum$: 
    a function $F$ is defined by $\WProd$ 
    if $F(y,x)=\prod^{\ell(y)}_{z=0}H(z,x)$ and by $\Sum$ from $H$ if $F(y,x)=\sum^{y}_{z=0} H(z,x)$.
Notably, in \cite{vollmer:96}, $\sP$ is characterized by the closure of $\Sum$ not only under $\FP$, but also under a smaller class (in turn defined using $\WProd$)~\cite{wagner:86}.

The class $\bff{FPSPACE}$ was characterized by Clote in~\cite{clote:99} relying on a scheme called $k$-bounded recursion, very similar to $k$-$\mathsf{BRN}$: given a constant $k\ge 4$, a function $F$ is said to be defined by $k$-$\mathsf{BR}$ from $G$ and $H$ if $F(\eps,x)=G(x)$ and $F(y+1,x)=H(y,x,F(y,x))$ where $F$ takes only values in $\{0,\dots, k\}$.
\section{Formulation of {\sf B}}\label{app:formulation_B}
\begin{definition}\label{def:B_all}
${\sf B}$ is the theory defined over $\cal{L}$, based on classical logic with equality, containing further the universal closures of the following non-logical axioms:

\,

\noindent I. Combinatory algebra:\\
(1) $\ms{k} x y = x$,\\
(2) $\ms{s} xyz = x z (y z)$

\,

\noindent II. Words, binary successor and predecessor:\\
(3) $\ms{W}(\epsilon)$,\\
(4) $(\forall x\in\ms{W})(\ms{W}(\ms{s_0} x) \wedge\ms{W}(\ms{s_1} x))$,\\
(5) $\ms{s_0} x \neq \ms{s_1} x\,\,\wedge\,\, \ms{s_0} x \neq\epsilon\,\,\wedge\,\,\ms{s_1} x \neq \epsilon$,\\
(6) $\ms{p}_\ms{W} : \ms{W}\to \ms{W}$,\\
(7) $\ms{p}_\ms{W}\epsilon = \epsilon$,\\
(8) $\ms{W}(x)\to \ms{p}_\ms{W}(\ms{s_0} x)= x\,\wedge\, \ms{p}_\ms{W}(\ms{s_1} x )= x$,\\
(9) $\ms{W}(x) \to x = \epsilon\vee  x= \ms{s_0}(\ms{p_w} x) \vee x = \ms{s_1}(\ms{p_w} x)$

\,

\noindent III. Lexicographic successor and predecessor:\footnote{As an example, successive applications of $\ms{s_\ell}$ to $\epsilon$ yield the sequence $\epsilon,0,1,00,01,10,11,000,\cdots$.}\\
(10) $\ms{s_\ell}:\ms{W}\to\ms{W}\wedge \ms{s_\ell}\epsilon = 0$,\\
(11) $x \in\ms{W}\to \ms{s_\ell}(\ms{s_0} x) = \ms{s_1}x \wedge \ms{s_\ell}(\ms{s_1} x) = \ms{s_0} (\ms{s_\ell} x)$,\\
(12) $\ms{p_\ell} : \ms{W} \to \ms{W} \wedge \ms{p_\ell}\ms{\epsilon} = \ms{\epsilon}$,\\
(13) $x \in\ms{W}\to \ms{p_\ell}(\ms{s_\ell}x) = x$,\\
(14) $x \in\ms{W} \wedge x \neq\epsilon \to \ms{s_\ell}(\ms{p_\ell} x) = x$

\,

\noindent IV. Conditional on words:\\
(15) $\ms{c_W} \epsilon y z w = y$,\\
(16) $\ms{W}(x)\to\ms{c_W} {(}\ms{s_0} x{)} y z w = z\,\,\wedge\,\,\ms{c_W} {(}\ms{s_1} x{)} y z w = w$

\,

\noindent V. Word concatenation and multiplication (using infix notation):\\
(17) $\ms{*}:\ms{W}^2\to \ms{W}$,\\
(18) $\ms{W}(x)\to x \ms{*} \epsilon = x$,\\ 
(19) $\ms{W}(x)\wedge \ms{W}(y)\to x \ms{*} (\ms{s_0} y) = \ms{s_0}(x\ms{*}y)\,\,\wedge\,\, x \ms{*} (\ms{s_1} y) = \ms{s_1} (x \ms{*} y)$,\\
(20) $\ms{\times} :\ms{W}^2\to \ms{W}$,\\
(21) $\ms{W}(x) \to x \ms{\times} \epsilon = \epsilon$,\\
(22) $\ms{W}(x)\wedge \ms{W}(y)\to x \ms{\times} (\ms{s_0} y) = (x\ms{\times} y)\ms{*} x\,\,\wedge\,\,  x \ms{\times}(\ms{s_1} y) =  (x\ms{\times}  y) \ms{*} x$

\,

\noindent VI. Bit\\
(23) $\ms{bit}:\ms{W}^2\to\ms{W}$,\\
(24) $\ms{W}(x) \to \ms{bit}x\eps =\eps$,\\
(25) $\ms{W}(x) \to \ms{bit}\eps x=\ms{c_W}x\eps01$,\\
(26) $\ms{W}(y)\wedge \ms{W}(x) \to \ms{bit}(\ms{s_\ell}y)(\ms{s_0}x) = \ms{bit}y x\wedge \ms{bit}(\ms{s_\ell}y)(\ms{s_1}x) = \ms{bit}y x$

\,

\noindent VII. Erasing function\\
(27) $\ms{e}: \ms{W}\to\ms{W}$,\\
(28) $\ms{e}\eps = \eps$,\\
(29) $\ms{W}(x)\to \ms{e}(\ms{s_0}x)= \ms{e}x$,\\
(30) $\ms{W}(x) \to \ms{e}(\ms{s_1}x) = \ms{s_1}(x)$

\,

\noindent VIII. Length function\\
(31) $\ms{len} : \ms{W}\to\ms{W}$,\\
(32) $\ms{len}(\ms{\eps}) = \eps$,\\
(33) $\ms{W}(x) \to \ms{len}(\ms{s_i} x) = \ms{s_\ell}(\ms{len} x)$

\,

\noindent IX. Extensionality:\\
(34) $(\forall x)(fx = gx)\to f = g$ \qquad (for $x$ not a free variable in $f$ or $g$)
\end{definition}
\section{Lower bound for \texorpdfstring{$\bff{FAC}^0$}{FAC0}}\label{app:alterations_AC0}
\begin{proposition}\label{prop:lower_bound_ac0}
$\bff{FAC}^0\subseteq\comp{LHT}$.
\end{proposition}
\begin{proof}
The initial functions of Eberhard's characterization of $\bff{FAC}^0$ \cite{eberhard:15} (which is similar to \cite{clote:99} but defined for the set $\bb{W}$) are axiomatized as total, so we only have to show that $\comp{LHT}$ is closed under {\sf CRN}.
Let $F$ be defined by $\mathsf{CRN}$ from two functions $G$ and $H$ already assumed to be provably total:
\begin{align*}
    &F(\ms{\epsilon},x) = G(x)&\\
    &F(yi,x) = S_{H(yi,x)}(F(y,x))&
\end{align*}
where $H(yi,x)\in\{0,1\}$. Let $g,h$ be $\cal{L}$-terms representing $G$ and $H$, whose totality can be proven in ${\sf LHT}$ by induction hypothesis.

Define the $\cal{L}$-term $f = \mathtt{rec}\,t$, where
\begin{align*}
    &t = \lambda fyx.\,\ms{c_\eps}y(gx)\big(\ms{c_W}(hyx)\,(\ms{s_0}(f(\ms{p_w}y)x))\,(\ms{s_0}(f(\ms{p_w}y)x))\,(\ms{s_1}(f(\ms{p_w}y)x))\big)&
\end{align*}
By Lemma \ref{lemm:fixpoint}, the following identities are provable in ${\sf LHT}$:
\begin{align*}
    &f\ms{\epsilon} x = t(\mathtt{rec}\,t)\ms{\epsilon} x = tf\ms{\epsilon} x = gx&\\
    &f(yi)x = tf(yi)x = \ms{c_W}(h(yi)x)\,(\ms{s_0}(fyx))\,(\ms{s_0}(fyx))\,(\ms{s_1}(fyx))&
\end{align*}
Thus, clearly $f$ represents the function $F$ in the sense of Definition \ref{def:provably_total_function}.

Now, we show that ${\sf LHT}$ proves the totality of $f$.
Let $A(y,z)$ be the atomic formula $fyx = z$.
Fix the parameter $x\in \ms{W}$. Since $f\ms{\epsilon} x = gx$, and $g$ is provably total by assumption, from $\ms{W}(x)$ it follows that $\ms{W}(gx)$, so $\exists z\in\ms{W}.A(\ms{\epsilon},z)$ holds.
Now, let $y\in\ms{W}$ and assume $A(y,z)$. 
Then, $f(yi)x = \ms{s_0}(fyx)$ or $f(yi)x = \ms{s_1}(fyx)$, so $A(yi,z0)\vee A(yi,z1)$ holds.
Thus, from $(\BSI)$, the totality of $f$ follows.
\end{proof}

It is simple to see how to extend the axiom $(\BSI)$ to allow concatenation with the empty word (by adding the disjunct $A(yi,z)$) or words of a size greater than $1$ (by adding further disjuncts).
\section{Upper bounds}\label{sec:upper_bounds_appendix}
We give more details for the upper bounds. First, the definition of realizability we work with is as follows:

\begin{definition}[\cite{strahm:03}]\label{def:realizability_appendix}
For a positive formula $A$ and a word $\rho\in\bb{W}$, the relation $\rhdd$ ($\rho$ ``realizes'' $A$) is inductively defined as follows:
\begin{align*}
    &\rho \rhdd \ms{W}(t) &\text{if}& 
    \qquad\mathcal{M}(\lambda\eta^+) \vDash t = \overline{\rho}\\
    &\rho \rhdd (t_1 = t_2) &\text{if}& 
    \qquad\rho = \epsilon \text{ and } \mathcal{M}(\lambda\eta^+) \vDash t_1 = t_2\\
    &\rho \rhdd (A\wedge B) &\text{if}& 
    \qquad\rho = \pair{\rho_0,\rho_1} \text{ and } \rho_0 \rhdd A \text{ and } \rho_1 \rhdd B\\
    &\rho \rhdd (A\vee B) &\text{if}& 
    \qquad\rho = \pair{i,\rho'} \text{ and either } i=0 \text{ and } \rho' \rhdd A \text{ or } i=1 \text{ and } \rho' \rhdd B\\
    &\rho \rhdd (\forall x)\,A(x) &\text{if}& 
    \qquad\rho \rhdd A(u) \text{ for a fresh variable } u\\
    &\rho \rhdd (\exists x)\,A(x) &\text{if}& 
    \qquad\rho \rhdd A(t) \text{ for some term } t \text{ that is free for } x \text{ in } A.
\end{align*}
\end{definition}

The rule corresponding to the axioms $(\SSI)$ and $(\SSI)^\vee$ is
\begin{prooftree}
    \AxiomC{
    $\begin{matrix}
    \Gamma\,\,\To\,\,\exists z\in\ms{W}. A(\epsilon,z),\Delta\\
    \Gamma,\ms{W}(y),A(y,z) \,\,\To\,\, A(y0,z),A(y0,\ms{s_\ell}z),\Delta\\
    \Gamma,\ms{W}(y),A(y,z) \,\,\To\,\, A(y1,z),A(y1,\ms{s_\ell}z),\Delta
    \end{matrix}$}
    \RightLabel{$R_{(\SSI)}/R_{(\SSI)}^\vee$}
    \UnaryInfC{$\Gamma,\ms{W}(s)\,\, \To\,\,\exists z\in\ms{W}\,A(s,z),\Delta$}
\end{prooftree}
with $A\in\cal{E}$ and $A\in\cal{E}^\vee$, respectively and with $z$ and $y$ free (and fresh) variables not occurring in $\Gamma$ or $\Delta$.
\subsection{Upper bound for \texorpdfstring{$\bff{FAC}^0$}{FAC0}}\label{sec:upper_bound_fac0}
\begin{proposition}\label{prop:upper_bound}
$\comp{LHT}\subseteq\bff{FAC}^0$.
\end{proposition}
\begin{proof}
Consider an instance of $(\BSI)$ and assume that the premises are realized by $G$, $H_0$ and $H_1$.
Let $\rho_A$ be the realizer for $A\in\cal{E}$ from Remark \ref{rmrk:easy_realizer}.
Define the following function by $\mathsf{CRN}$:
\begin{align*}
    &F'(\vec{\rho},\ms{\epsilon}) = ((G(\vec{\rho}))_1)_0 &\\
    &F'(\vec{\rho},\sigma i) =  S_{H_i(\vec{\rho},\sigma,\rho_A)_0}(F'(\vec{\rho},\sigma))&
\end{align*}
and let $F(\vec{\rho},\sigma)$ be defined as follows, on inputs $\vec{\rho},\sigma$: 
\begin{enumerate}
    \item Run Algorithm \ref{alg:side_formulas}. If it does not output $\eps$, return its output (a realizer for the side formulas)
    \item If $\sigma = \eps$, return $G(\vec{\rho})$
    \item Otherwise, return $\pair{0,\pair{F'(\vec{\rho},\sigma),\rho_A}}$
\end{enumerate}
The function $F$ is in $\bff{FAC}^0$ and realizes the sequent.
\end{proof}
\subsection{Upper bound for the counting hierarchy}\label{app:uppBCounting}

We give the upper bound for the counting hierarchy in more detail.

\begin{proposition}\label{prop:upper_bound_CH}
If the premises in an application of $(\SSIl)$ are realized by $\bff{FCH}$ functions, so is the conclusion.
\end{proposition}
\begin{proof}
Consider an application of $(\SSIl)$ and let $G$ and $H$ be the two $\bff{FCH}$ functions which realize the two premises of $(\SSIl)$.
We construct below an algorithm for a $\bff{FCH}$ function $F$ such that:
for every list of terms $\vec{s}\in\cal{L}$ and $\vec{\rho},\sigma\in\bb{W}$, if $\vec{\rho}$ are realizers of all the formulas in $\Gamma[\vec{s}]$ and $\sigma\rhdd\ms{W}(s)$, then $F(\vec{\rho},\sigma)\rhdd \exists z\in\ms{W}.\,A(s,z),\Delta[\vec{s}]$.\footnote{Recall that in the rule representing $(\SSIl)$, the variable $z$ does not occur in $\Gamma$ or $\Delta$.}

From Remark \ref{rmrk:easy_realizer}, let $\rho_A$ be a realizer for $A$.

\begin{algorithm}[H]
\SetArgSty{upshape}
\DontPrintSemicolon
\caption{Realizing $(\SSIl)$}
\textbf{Input} $\vec{\rho},\sigma$\;
\If{$\sigma = \eps$ or $(G(\vec{\rho}))_0\neq 0$}{
    \Return $G(\vec{\rho})$\;
}
\If{there is $\tau<_\ell \sigma$ such that $(H(\vec{\rho},\tau,\rho_A))_0\notin\{0,1\}$}{
    $\tau:=$ smallest word satisfying the above condition\;
    \Return $\pair{(H(\vec{\rho},\tau,\rho_A))_0-1,(H(\vec{\rho},\tau,\rho_A))_1}$\;
}
$n := \#\{\tau<_\ell\sigma\,:\,(H(\vec{\rho},\tau,\rho_A))_0=1\}$\;
\Return $\pair{0,\pair{((G(\vec{\rho}))_1)_0+n,\rho_A}}$
\label{alg:realizing_ch}
\end{algorithm}

First, about the complexity of the function.
Since $G$ and $H$ are in $\bff{FCH}$, which is closed under existential search, computing the ``if statements'' can be done in $\bff{FCH}$. Since $\bff{FCH}$ is also closed under bounded minimization, returning the condition after the second ``if statements'' can also be done in $\bff{FCH}$.
Finally, computing the number of times that $H$ realized the disjunct $A(\ms{s_\ell}y,\ms{s_\ell}z)$ is a particular case of bounded sum, under which $\bff{FCH}$ is also closed.
Thus, $F\in\bff{FCH}$.

\smallskip

Now, to see that $F$ realizes the given sequent, let $\vec{s}$ be a list of terms instantiating the free variables of $\Gamma$ and $\Delta$, let $\vec{\rho}$ be realizers for the formulas in $\Gamma[\vec{s}]$, and let $\sigma\rhdd\ms{W}(s)$, i.e.,~$s=\overline{\sigma}$. We distinguish three cases.

\emph{Case 1: $(G(\vec{\rho}))_0\neq 0$.}
Then $G(\vec{\rho})=\pair{j,\rho'}$, where $j\geq 1$ is the index of a formula in $\Delta[\vec{s}]$ realized by $\rho'$.
Since both the succedent of the first premise and the succedent of the conclusion consist of one main formula (at index $0$) followed by the formulas of $\Delta$, $F(\vec{\rho},\sigma)=G(\vec{\rho})$ realizes $\Delta[\vec{s}]$, and hence the succedent of the conclusion.

\emph{Case 2: $(G(\vec{\rho}))_0=0$ and there is $\tau<_\ell\sigma$ with $(H(\vec{\rho},\tau,\rho_A))_0\notin\{0,1\}$.}
Let $\tau$ be the smallest such word.
Since $G(\vec{\rho})=\pair{0,\pair{w,\rho'}}$ realizes $\exists z\in\ms{W}.A(\epsilon,z)$, there is a term $r_\epsilon$, with value $w=((G(\vec{\rho}))_1)_0$, such that $A(\overline{\epsilon},r_\epsilon)$ is true.
Since $\tau$ is the smallest word satisfying the condition, $(H(\vec{\rho},\nu,\rho_A))_0\in\{0,1\}$ holds for all $\nu<_\ell\tau$.
Hence, by induction on the lexicographic position of $\nu\leq_\ell\tau$, we can define terms $r_\nu$ such that $A(\overline{\nu},r_\nu)$ is true: given $r_\nu$, the inputs $\vec{\rho}$, $\nu$ and $\rho_A$ realize $\Gamma[\vec{s}]$, $\ms{W}(\overline{\nu})$ and $A(\overline{\nu},r_\nu)$, respectively (recall that, since $A\in\cal{E}$, $\rho_A$ realizes $A(\overline{\nu},r_\nu)$ whenever it is true, Remark~\ref{rmrk:easy_realizer}).
Thus, $H(\vec{\rho},\nu,\rho_A)$ realizes one of $A(\ms{s_\ell}\overline{\nu},r_\nu)$, $A(\ms{s_\ell}\overline{\nu},\ms{s_\ell}r_\nu)$ or a formula in $\Delta[\vec{s}]$, and by minimality of $\tau$ it is one of the first two.
Accordingly, we set $r_{\ms{s_\ell}\nu}:=r_\nu$ if $(H(\vec{\rho},\nu,\rho_A))_0=0$, and $r_{\ms{s_\ell}\nu}:=\ms{s_\ell}r_\nu$ otherwise.
In particular, $A(\overline{\tau},r_\tau)$ is true, so that $H(\vec{\rho},\tau,\rho_A)=\pair{j,\rho''}$, with $j\notin\{0,1\}$, is called on realizers of $\Gamma[\vec{s}]$, $\ms{W}(\overline{\tau})$ and $A(\overline{\tau},r_\tau)$, and therefore realizes a formula of the succedent of the second premise: since $j\geq 2$, this is the $(j-2)$-th formula of $\Delta[\vec{s}]$, realized by $\rho''$ (recall that $z$ does not occur in $\Delta$, so the choice of $r_\tau$ for $z$ is immaterial).
In the succedent of the conclusion, the same formula has index $j-1$, whence $F(\vec{\rho},\sigma)=\pair{j-1,\rho''}$ realizes $\Delta[\vec{s}]$.

\emph{Case 3: $(G(\vec{\rho}))_0=0$ and $(H(\vec{\rho},\tau,\rho_A))_0\in\{0,1\}$ for all $\tau<_\ell\sigma$.}
Then the construction of Case 2 yields terms $r_\nu$ with $A(\overline{\nu},r_\nu)$ true for all $\nu\leq_\ell\sigma$.
By construction, $r_\sigma$ is obtained from $r_\epsilon$ by applying $\ms{s_\ell}$ once for every $\nu<_\ell\sigma$ with $(H(\vec{\rho},\nu,\rho_A))_0=1$, i.e.,~$r_\sigma=\ms{s_\ell}^{\,n}r_\epsilon$ for $n=\#\{\tau<_\ell\sigma\,:\,(H(\vec{\rho},\tau,\rho_A))_0=1\}$.
Since $\ms{s_\ell}:\ms{W}\to\ms{W}$ and $r_\epsilon$ has value $w=((G(\vec{\rho}))_1)_0$, the term $r_\sigma$ has value $w+n$ (computed by lexicographic addition) and satisfies $\ms{W}(r_\sigma)$.
Moreover, $A(\overline{\sigma},r_\sigma)$ is true, so $\rho_A\rhdd A(s,r_\sigma)$.
Therefore $w+n\rhdd\ms{W}(r_\sigma)$ and, taking $r_\sigma$ as witness for the existential quantifier,
$F(\vec{\rho},\sigma)=\pair{0,\pair{w+n,\rho_A}}\rhdd \exists z\in\ms{W}.\,A(s,z),\Delta[\vec{s}]$. 
\end{proof}

\section{Proofs from Section~\ref{sec:separatin_oracle}}\label{app:oracle_in_theory}

\begin{proof}[Proof of Prop.~\ref{prop:oracle_in_theory}]
Each axiom and rule of {\sf ALT} can be realized by a $\bff{FNC}^1$ function and thus, in particular, by a $\bff{FP}$ function.
The extra axiom is $\ms{W}(x)\To\ms{W}(1_A(x))$, which is realized by the $\bff{FP}^A$ function $F(\rho) = \pair{0,1_A(\rho)}$.
Now, the statement follows by noticing that the realizability theorems \textit{relativize}, as they consist of simple simulations of the realizing functions given by induction. As an example, consider the cut rule (see e.g.~\cite{strahm:03,eberhard:15} 
on the other cases):
\begin{prooftree}
    \AxiomC{$\Gamma\Rightarrow A,\Delta$}
    \AxiomC{$\Gamma,A\Rightarrow \Delta$}
    \RightLabel{(cut)}
    \BinaryInfC{$\Gamma\To\Delta$}
\end{prooftree}
where $A$ is  positive. Let the premises be realized by (resp.) $G$ and $H$ in $\bff{FP}^A$, and define:
\begin{align*}
    &F(\vec{\rho}) = 
    \begin{cases}
        \langle (G(\vec{\rho}))_0-1, (G(\vec{\rho}))_1\rangle & \text{ if } (G(\vec{\rho}))_0\neq 0\\
        H(\vec{\rho},(G(\vec{\rho}))_1), & \text{ otherwise}
    \end{cases}&
\end{align*}
Then, $F$ realizes $\Gamma\To\Delta$ and is clearly in $\bff{FP}^A$ as well.
\end{proof}

\end{document}